\documentclass{nlaauth}
\usepackage{graphicx}
\usepackage{moreverb}
\usepackage{multirow}
\usepackage{hyperref}
\usepackage{setspace}
\usepackage[ruled,titlenotnumbered,nofillcomment]{algorithm2e}
\usepackage{subfigure}
\usepackage{color}

\newcommand\BibTeX{{\rmfamily B\kern-.05em \textsc{i\kern-.025em b}\kern-.08em
T\kern-.1667em\lower.7ex\hbox{E}\kern-.125emX}}

\graphicspath{{./}}

\usepackage{boxedminipage}

\newcommand{\mr}[2]{\multirow{#1}{*}{#2}}
\newcommand{\mc}[3]{\multicolumn{#1}{#2}{#3}}

\renewcommand{\div}{\nabla\cdot\,}
\newcommand{\grad}{\ensuremath{\nabla}}

\newcommand{\bfa}{{\bf a}}
\newcommand{\bfe}{{\bf e}}

\newcommand{\bfx}{{\bf x}}

\newcommand{\bfu}{{\bf u}}
\newcommand{\bfq}{{\bf q}}

\newcommand{\bfn}{{\bf n}}

\newcommand{\bfr}{{\bf r}}

\newcommand{\bfb}{{\bf b}}

\begin{document}


\runningheads{Eran Treister and Eldad Haber}
{A multigrid solver for the Helmholtz equation with a point source}

\title{A multigrid solver to the Helmholtz equation with a point source based on travel time and amplitude}

\author{
 Eran Treister\affil{1} \corrauth\ and
 Eldad Haber\affil{2}
}

\address{\centering \affilnum{1}\ Department of Computer Science, Ben-Gurion University of the Negev, Beer Sheva, Israel. ({\tt erant@cs.bgu.ac.il}).\\
\affilnum{2}\ Department of Earth and Ocean Sciences, The University of British Columbia, Vancouver, BC, Canada. ({\tt ehaber@eos.ubc.ca}).}

\corraddr{{\tt erant@cs.bgu.ac.il}}



\begin{abstract}
The Helmholtz equation arises when modeling wave propagation in the frequency domain. The equation is discretized as an indefinite linear system, which is difficult to solve at high wave numbers.
In many applications, the solution of the Helmholtz equation is required for a point source. In this case, it is possible to reformulate the equation as two separate equations: one for the travel time of the wave and one for its amplitude. The travel time is obtained by a solution of the factored eikonal equation, and the amplitude is obtained by solving a complex-valued advection-diffusion-reaction (ADR) equation. The reformulated equation is equivalent to the original Helmholtz equation, and the differences between the numerical solutions of these equations arise only from discretization errors. We develop an efficient multigrid solver for obtaining the amplitude given the travel time, which can be efficiently computed. This approach is advantageous because the amplitude is typically smooth in this case, and hence, more suitable for multigrid solvers than the standard Helmholtz discretization. We demonstrate that our second order ADR discretization is more accurate than the standard second order discretization at high wave numbers, as long as there are no reflections or caustics. Moreover, we show that using our approach, the problem can be solved more efficiently than using the common shifted Laplacian multigrid approach.
\end{abstract}


\maketitle

\section{Introduction}

The acoustic Helmholtz equation is used to model the propagation of a wave within a heterogeneous medium. Assuming constant density, the equation is given by
\begin{equation}\label{eq:Helmholtz}
\Delta u + \omega^2 \kappa^2(\vec{x})u = q(\vec{x}), \quad \vec{x}\in\Omega,
\end{equation}
where $u(\vec{x})$ is the pressure wave function in the frequency domain, $\omega=2\pi f$ is the angular frequency and $\kappa(\vec{x})$ is the ``slowness'' of the medium---the inverse of its velocity. The right-hand-side $q(\vec x)$ is used to incorporate sources into the equation. In this work we consider the case where $q(\vec{x}) = \delta(\vec{x} - \vec{x}_0)$, which models a point source at location $x_0$ ($\delta(\cdot)$ is the Dirac delta function). The Helmholtz equation with a point source is common in geophysical applications, e.g. seismic modeling and full-waveform inversion \cite{pratt1999,EpanomeritakisAkcelikGhattasBielak2008,virieux2009overview,krebs09ffw,biondi2014simultaneous,metivier2013full,JointEikFWI17}.

The Helmholtz equation is accompanied with boundary conditions, which can be Neumann or Dirichlet for example. In many cases the equation is involved with absorbing boundary conditions that mimic the propagation of a wave in an open domain. One option for this is the Sommerfeld boundary condition
\begin{equation}\label{eq:sommerfeld}
\bfn\cdot \nabla u - i\omega\kappa u = 0.
\end{equation}
A more effective way to absorb the waves is by using a boundary layer. This can be achieved by either a perfectly matched layer (PML) \cite{singer2004perfectly} or an absorbing boundary layer \cite{engquist1979radiation,liao1996multifrequency,erlangga2006novel}. To implement the latter layer, for example, we add an attenuation term to Eq. \eqref{eq:Helmholtz}:
\begin{equation*}
\Delta u + \omega^2 \kappa^2(\vec{x})u - i\omega\gamma\kappa^2(\vec{x})u = q(\vec{x}), \quad \vec{x}\in\Omega,
\end{equation*}
where $\gamma(\vec{x})\geq 0$ is a function that quadratically goes from 0 to $\omega$ towards the boundaries of the domain, which attenuates the waveform towards the these boundaries \cite{erlangga2006novel}. The thickness of this layer is usually chosen to be about one wavelength. The same $\gamma$ parameter can be used to impose attenuation all over the domain, but is quite small in most realistic scenarios.  To ease the derivations in this paper, we henceforth ignore the boundary layer and attenuation, i.e., assume that $\gamma=0$, and focus on the equation \eqref{eq:Helmholtz}.

We are mostly interested in problems where the frequency $\omega$ (or the wavenumber $\kappa\omega$) is high. In this case, the resulting linear system which arises from the discretization of \eqref{eq:Helmholtz} is highly indefinite. While 2D solutions can be obtained using direct methods, solving the discretized equation in 3D is challenging. That is because the discretization of the problem requires a very fine mesh and a large number of unknowns, in addition to the indefiniteness of the associated matrix \cite{calandra2013improved,poulson2013parallel,haber2011fast}.

In recent years, there has been a great effort to develop efficient solvers for systems arising from \eqref{eq:Helmholtz}, using several different approaches to tackle the problem. One of the most common approaches is the shifted Laplacian multigrid preconditioner \cite{erlangga2006novel,oosterlee2010shifted,airaksinen2007algebraic,erlangga2008multilevel,calandra2013improved,cools2014new,tsuji2015augmented,Tobias2017,ganesh2017efficient}, which modifies the equation by adding complex values to the diagonal of the matrix. The modified system is then solved using a multigrid method, and is used as a preconditioner for the non-shifted system to obtain the solution of the problem. Another recent approach was recently proposed in \cite{engquist2011sweeping,poulson2013parallel} for solving \eqref{eq:Helmholtz} in 2D and 3D respectively. The approach can be viewed as a domain decomposition method with particular boundary conditions, and can be effective in terms of iterations, but requires a large setup time and storage. This can impose challenges if the solution of \eqref{eq:Helmholtz} is required for multiple frequencies. Other iterative approaches include \cite{van2012preconditioning,gordon2013robust}, and \cite{brandt1997wave,olson2010smoothed,haber2011fast,livshits2014scalable} which are multigrid-based.

In this paper we develop a new approach for solving the Helmholtz equation based on \cite{haber2011fast}. Rather than solve the discrete
\eqref{eq:Helmholtz}, we reformulate the problem by using the Rytov decomposition of the solution\footnote{The Rytov decomposition is usually given by $u(\vec x) = a(\vec x)\exp(i\omega\tau(\vec x))$. Here we use its complex conjugate, which is equivalent to using the standard decomposition.}
\begin{equation}\label{eq:new_u}
u(\vec x) = a(\vec x)\exp(-i\omega\tau(\vec x)).
\end{equation}
Here, the waveform $u(\vec x)$ is decomposed into an amplitude $a(\vec{x})$, which changes slowly, and a phase that oscillates, involving $\tau(\vec{x})$.
To reformulate \eqref{eq:Helmholtz} according to \eqref{eq:new_u}, we first use the chain rule to obtain
\begin{equation*}
\begin{array}{lcl}
\nabla u    & = & (\nabla a - i\omega \nabla\tau)\exp(-i\omega\tau),\\
\Delta u & = & (\Delta a - 2i\omega \nabla\tau \cdot \nabla a - i\omega\Delta\tau - \omega^2a|\nabla\tau|^2)\exp(i\omega\tau).
\end{array}
\end{equation*}
Then, after plugging $\Delta u$ into \eqref{eq:Helmholtz} and multiplying the equation by $\exp(i\omega\tau)$, we get the following complex-valued advection-diffusion-reaction (ADR) equation:
\begin{equation}\label{eq:ADR}
\Delta a - 2i\omega\nabla\tau \cdot \nabla a  -i\omega(\Delta\tau)a  - \omega^2(|\nabla\tau|^2-\kappa(\vec x)^2)a = \hat q(\vec x),
\end{equation}
where $\hat q(\vec x) = q(\vec x)\exp(i\omega\tau)$. This equation is also equivalent to
\begin{equation}\label{eq:ADR2}
\Delta a - i\omega\nabla\tau \cdot \nabla a  - i\omega \nabla\cdot(\nabla\tau a) - \omega^2(|\nabla\tau|^2-\kappa(\vec x)^2)a =  \hat q(\vec x),
\end{equation}
which does not contain the term $\Delta\tau$.

We note that the function $a(\vec{x})$ does not fully correspond to a real amplitude, as usually assumed when using the Rytov decomposition. For example, $a(\vec{x})$ is not necessarily positive or even real valued. In the process above we artificially doubled the unknowns of Eq. \eqref{eq:Helmholtz} using the Rytov decomposition. Hence, for any given $\tau(\vec{x})$ and frequency $\omega$, the function $a(\vec{x})$ is unique according to \eqref{eq:new_u}, and the resulting Eq. \eqref{eq:ADR}-\eqref{eq:ADR2} are equivalent to \eqref{eq:Helmholtz}. Their numerical solutions using \eqref{eq:new_u} are equivalent to the numerical solution of \eqref{eq:Helmholtz} up to discretization and roundoff errors \emph{only}, given equivalent boundary conditions (to be discussed later). For example, if we choose $\tau=0$, then we retain the equation \eqref{eq:Helmholtz}, and $a(\vec{x}) = u(\vec{x})$.
The work in \cite{haber2011fast} aims to get a multigrid preconditioner for \eqref{eq:Helmholtz}, and chooses $\tau(\vec{x})$ to be a plane, so that the problem \eqref{eq:ADR} becomes positive definite and easy to solve. In this work we suggest new discretizations for the Helmholtz problem based on \eqref{eq:ADR} or \eqref{eq:ADR2}. Our first goal is to choose $\tau(\vec x)$ such that the amplitude $a(\vec{x})$ is as smooth as possible, so we get small numerical errors when we discretize \eqref{eq:ADR} or \eqref{eq:ADR2}. Our second goal is to get a linear system which can be solved efficiently by multigrid methods.

\begin{figure}
\begin{center}
	\newcommand{\image}[1]{\includegraphics[width=0.24\linewidth]{#1}}
  \subfigure[The waveform $u(\vec{x})$]{\image{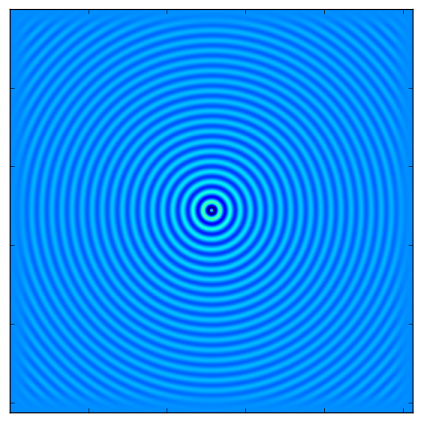}}
  \subfigure[The travel time $\tau(\vec x)$]{\image{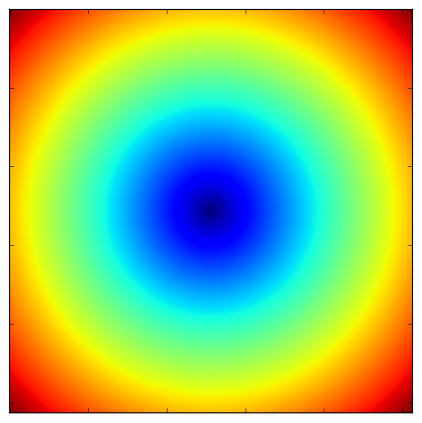}}
\subfigure[The amplitude $a(\vec x)$]{\image{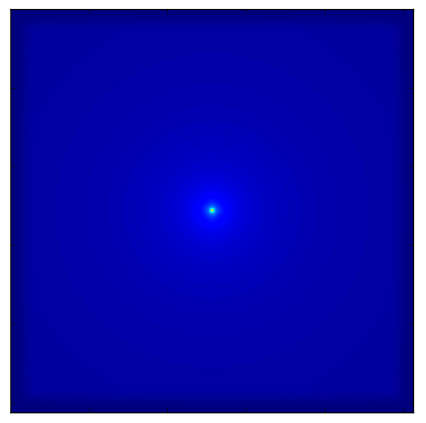}}
\subfigure[The phase $\exp(-i\omega\tau)$ ]{\image{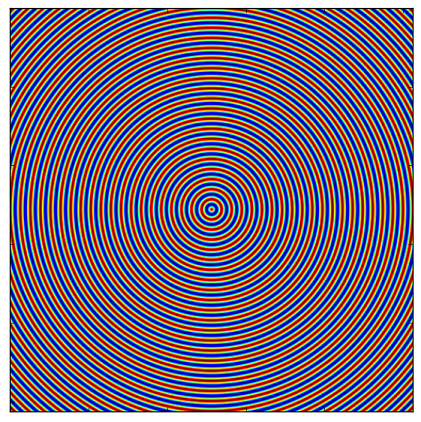}}
\end{center}
\caption{The solution of the Helmholtz equation with homogenous media for a point source with absorbing boundary conditions. The oscilatory waveform is given by $u = a\exp(-i\omega\tau)$, while the $a(\vec{x})$ and $\tau(\vec{x})$ that are calculated numerically are smooth.}
\label{fig:constExample}
\end{figure}

To fulfil our first goal, we aim to capture most of the oscillatory behaviour of the solution $u(\vec{x})$ by choosing $\tau(\vec{x})$ appropriately, so that the corresponding amplitude $a(\vec{x})$ is smooth. As motivation, consider the solution of \eqref{eq:Helmholtz} with a point source for a constant medium $\kappa = 1$. In 3D, the analytical solution is $u(\vec x) = \frac{1}{4\pi r}\exp(-i\omega r)$, where $r = \|\vec x - \vec x_0\|_2$ is the euclidian distance from the source $\vec x_0$. In this case, if we choose $\tau(\vec x) = r$ we get $a(\vec x) = \frac{1}{4\pi r}$, which is a rather smooth function (except at $x_0$). Figure \ref{fig:constExample} demonstrates this case in 2D. For a heterogenous medium, a similar result can be obtained by solving the eikonal equation
\begin{equation}\label{eq:eikonal}
|\nabla\tau|^2 = \kappa(\vec{x})^2,
\end{equation}
where $|\cdot|$ is the Euclidean norm. This is an advection equation that requires a known initial value at some sub-region, and in the context of wave propagation, the solution $\tau$ has the meaning of travel time, or first arrival time, of the wave. In this work we consider
the wave propagation from a point source at location $\vec{x}_0$, for which the travel time is 0, and hence $\tau(\vec{x}_{0}) = 0$. By choosing $\tau$ according to \eqref{eq:eikonal}, we eliminate the last term in the left-hand-side of \eqref{eq:ADR}.

The idea of using travel time and amplitude for modeling wave propagation from a point source was studied recently in a different approach than in this paper. The Rytov decomposition \eqref{eq:new_u} is used to get a geometrical-optics $O(\frac{1}{\omega})$ ansatz for the solution of the Helmholtz equation in the high frequency regime \cite{leung2007eulerian,luo2011factored,luo2012higher,luo2014fast}. There, similarly to our approach, the eikonal equation is used to eliminate the $\omega^2$ term in \eqref{eq:ADR} yielding the travel time, and the amplitude is obtained by eliminating the $\omega$ term in \eqref{eq:ADR}, by solving the real-valued transport equation
\begin{equation}\label{eq:transport}
\nabla\tau\cdot\nabla a + \frac{1}{2}(\Delta\tau)a = \frac{1}{2}(\nabla\tau\cdot\nabla a + \div ((\grad\tau) a)) = 0.
\end{equation}
The resulting approximation includes only the first arrival information of the wave propagation, and since \eqref{eq:transport} does not depend on $\omega$, this approximation aims to be valid for multiple frequencies. Our approach is different in the way that the amplitude is defined, according to \eqref{eq:ADR} instead of \eqref{eq:transport}. This way, the amplitude is specific for a given frequency and contains all the information of the wave propagation, e.g. it includes reflections and interferences.

To compute the phase for \eqref{eq:new_u}, the eikonal equation
is solved. This equation is nonlinear, and hence may have many solutions. Out of these, the solution that corresponds to the first arrival time can be computed efficiently \cite{crandall1983viscosity,rouy1992viscosity}.  One of the most effective ways to compute it is by Fast Marching methods \cite{tsitsiklis1995efficient,sethian1996fast,sethian1999fast}, which solve \eqref{eq:eikonal} directly using first or second order schemes in ${\cal O}(n\log n)$ operations, based on the monotonicity of the solution along the characteristics. Alternatively, \eqref{eq:eikonal} can be solved iteratively by Fast Sweeping methods with first or higher order of accuracy \cite{tsai2003fast, zhao2005fast,zhang2006high}. The equation \eqref{eq:eikonal} can also be solved using a Lax-Friedrichs scheme \cite{kao2004lax,qian2007fast}, which involves adding artificial viscosity to \eqref{eq:eikonal}.

However, the methods mentioned above are not suitable for solving \eqref{eq:eikonal} for our purpose. To get an accurate solution for the amplitude in \eqref{eq:transport}, the numerical approximation for $\tau(\vec{x})$ has to be very accurate \cite{luo2012higher}. Because the analytical $\tau(\vec{x})$ is non-smooth at the point source, the numerical solution of Eq. \eqref{eq:eikonal} is polluted with errors when it is computed using the aforementioned standard finite-difference methods \cite{fomel2009fast}. To overcome this, \cite{luo2011factored,luo2012higher} use the factored version of the eikonal equation which was originally suggested in \cite{pica1997fast}. The new equation is obtained by setting $\tau = \tau_0\tau_1$ in \eqref{eq:eikonal}, where the function $\tau_0(\vec{x})$ is known and its derivatives are computed analytically. Using the chain rule we get the factored eikonal equation for $\tau_1$
\begin{equation}\label{eq:factoredEikonal}
|\tau_0\nabla\tau_1 + \tau_1\nabla\tau_0|^2 = \kappa(\vec{x})^2.
\end{equation}
The most common choice for $\tau_0$ is the distance function from the point source, i.e., $\tau_{0} = \|\vec{x}-\vec{x}_0\|_2$. That is the analytical solution for \eqref{eq:eikonal} in the case where $\kappa(x)=1$. The function $\tau_0$ is non-smooth at the location of the source, but the factor $\tau_1$ which needs to be computed numerically is expected to be very smooth at the surrounding of the source. Similarly to it original version, Eq. \eqref{eq:factoredEikonal} can be solved directly by the Fast Marching method \cite{TH2016}, or iteratively by the Fast Sweeping methods with first order accuracy \cite{fomel2009fast,luo2012fast,LouQianBurridge2014}, or by a Lax-Friedrichs scheme up to third order of accuracy \cite{luo2012higher,luo2011factored,LouQianBurridge2014}. The works \cite{LouQianBurridge2014,noble2014accurate} suggest hybrid schemes where the factored eikonal equation is solved at the neighborhood of the source, and the standard eikonal equation is solved in the rest of the domain.

Similarly to \cite{luo2012higher}, in this work we use the factored eikonal equation \eqref{eq:factoredEikonal} to get an accurate solution for the Helmholtz equation based on \eqref{eq:ADR}-\eqref{eq:ADR2}. We apply the Fast Marching method suggested in \cite{TH2016} for solving \eqref{eq:factoredEikonal}, but in principle all the methods mentioned earlier for solving the factored eikonal equation can be used in our approach. Note that the equations for the amplitude that include $\Delta\tau$ now include $\Delta\tau_1$ in the factored case. That is a numerical approximation of the Laplacian of the factor $\tau_1$, which may be non-smooth in areas away from the source due to discontinuities in $\kappa$ or due to caustics. $\nabla\tau_1$ is not continuous but is bounded because of \eqref{eq:factoredEikonal}. Approximating the second derivative of a non-smooth function may yield high numerical errors. For this reason, third order Lax Friedrichs scheme, which adds smoothness to $\tau_1$, is used in \cite{luo2012higher,luo2011factored}. However, both the equations \eqref{eq:transport} and \eqref{eq:ADR2} show that the problems can be formulated without $\Delta\tau_1$, and there is no real need to approximate it numerically.

\bigskip

To summarize, in this work we reformulate the Helmholtz equation using ideas from \cite{haber2011fast,luo2011factored,luo2012higher,TH2016} to allow a more efficient numerical solution  of the equation for a point source. We aim that the majority of the solution is represented by smooth functions, which have physical meanings of amplitude and travel time. Similarly to \cite{haber2011fast}, we use the full ADR equation \eqref{eq:ADR} or \eqref{eq:ADR2} for the amplitude and solve it by multigrid methods. However, instead of choosing the travel time $\tau$ as a plane, we solve \eqref{eq:factoredEikonal} to find it, so that the amplitude $a(x)$ is smooth. The smoothness of $a(x)$ contributes to the efficiency of multigrid methods for solving \eqref{eq:ADR}, as smooth functions can be approximated well on coarser grids. The rest of the paper is organized as follows: in Section \ref{sec:discretization} we present the options that we consider for discretizing the ADR equation, and examine the similarity between the discretized Helmholtz and ADR equations. Then, in Section \ref{sec:MG} we present the multigrid methods that we use to solve the ADR and Helmholtz equations. Finally, in section \ref{sec:results} we present numerical results, that first compare the accuracy of the two approaches for the Helmholtz problem, and then compare the computational effort needed to solve the equations using multigrid.

\section{The discretization of the reformulated advection-diffusion-reaction equation}\label{sec:discretization}

The Helmholtz equation \eqref{eq:Helmholtz} is usually discretized using the finite difference method on a regular mesh. The standard approach involves a second order scheme, resulting in a five and seven point stencils for two and three dimensions, respectively. We denote the corresponding linear system by
\begin{equation}\label{eq:HelmholtzLinSystem}
H\bfu = \bfq.
\end{equation}
We note that using this discretization, one has to use a mesh that is fine enough, having at least 10-15 grid-point per wavelength \cite{erlangga2006novel,haber2011fast,calandra2013improved,poulson2013parallel}, otherwise the numerical solution is polluted by dispersion errors. This requirement leads to rather large matrices, and therefore, higher order discretizations were developed to minimize the numerical dispersion phenomenon and require fewer grid points per wavelength on the expense of larger stencils \cite{singer1998high,singer2006sixth,operto20073d,turkel2013compact}. In this work we discretize \eqref{eq:ADR}-\eqref{eq:ADR2} using second order stencils as in \cite{haber2011fast}, but note that extensions to higher order stencils can be obtained as well. In particular, we consider two types of discretizations for the ADR equations, which are different in the scheme used for the advection operator in \eqref{eq:ADR}. The first discretization includes a central difference scheme for the advection operator. We denote the resulting linear system
\begin{equation}\label{eq:ADRLinSystemCentral}
\hat H_{cen}\bfa = \hat\bfq.
\end{equation}
In the second discretization we use a second order upwind scheme for the advection term, and denote the resulting linear system
\begin{equation}\label{eq:ADRLinSystemUpwind}
\hat H_{2up}\bfa = \hat\bfq.
\end{equation}

We now present how we define \eqref{eq:ADRLinSystemCentral}-\eqref{eq:ADRLinSystemUpwind}, and later show a relation between the matrices $H$ and $\hat H_{cen}$. We present all the finite difference operators in 1D, and extensions to 2D and 3D are straightforward.

\subsection{The coefficients of the equation}
First, we define the coefficients of Eq. \eqref{eq:ADR}-\eqref{eq:ADR2}, involving the travel time $\tau$. For this, we assume that the travel time factor $\tau_1$ is calculated by Fast Marching \cite{TH2016} in second order of accuracy on the same mesh. Using the chain rule, we define
\begin{equation}\label{eq:travel_time_coeff}
\nabla\tau = \tau_0\nabla\tau_1 + \tau_1\nabla\tau_0 \quad \mbox{ and }\quad\Delta\tau = \tau_1\Delta\tau_0 + 2\nabla\tau_0\cdot\nabla\tau_1 + \tau_0\Delta\tau_1,
\end{equation}
where $\tau_0$, $\nabla\tau_0$, and $\Delta\tau_0$ are the known analytic solution of \eqref{eq:eikonal} for a constant medium and its analytic derivatives. To approximate $\nabla\tau_1$ in \eqref{eq:travel_time_coeff}, we use the second order central difference operator
$$
\left(\frac{\partial\tau_1}{\partial x}\right)_j \approx \frac{(\tau_1)_{j+1} - (\tau_1)_{j-1}}{2h},
$$
while reverting to first order operators on the domain boundaries. As an alternative, we may use the discrete gradients that where used to calculate $\tau_1$ in the factored Fast Marching algorithm. To approximate $\Delta\tau_1$ in \eqref{eq:travel_time_coeff} we use the standard second order central difference operator, and use the one sided second order second derivative operator on the boundaries, as there are no natural boundary conditions to $\tau$.

\subsection{The discretization of the ADR equation}
Once the coefficients of \eqref{eq:ADR}-\eqref{eq:ADR2} are known, we discretize the equations using second order finite difference operators. The Laplacian operator is discretized using a standard second order central difference operator, just as in the second order discretization of \eqref{eq:Helmholtz}. As noted before, for the advection term $\nabla\tau\cdot\nabla a$ we have two options:
\begin{enumerate}
\item Using a second order central difference:
\begin{equation}\label{eq:londDiff}
\left(\frac{\partial \tau}{\partial x}\cdot\frac{\partial a}{\partial x}\right)_j \approx \left(\frac{\partial \tau}{\partial x}\right)_j\cdot \frac{a_{j+1} - a_{j-1}}{2h}.
\end{equation}
This operator is used in \eqref{eq:ADRLinSystemCentral}
\item Using a second order upwind scheme:
\begin{equation}\label{eq:londUpwindDiff}
\left(\frac{\partial \tau}{\partial x}\cdot\frac{\partial a}{\partial x}\right)_j \approx \left\{
\begin{array}{cc}
\left(\frac{\partial \tau}{\partial x}\right)_j\cdot \frac{3a_{j} - 4a_{j-1} + a_{j-2}}{2h} & \mbox{if } \left(\frac{\partial \tau}{\partial x}\right)_j > 0\\
\left(\frac{\partial \tau}{\partial x}\right)_j\cdot \frac{-3a_{j} + 4a_{j+1} - a_{j+2}}{2h} & \mbox{if } \left(\frac{\partial \tau}{\partial x}\right)_j < 0
\end{array}\right..
\end{equation}
This operator is used in \eqref{eq:ADRLinSystemUpwind}.
\end{enumerate}
\noindent   One important observation should be made regarding \eqref{eq:londUpwindDiff}: at the immediate neighborhood of the source point (distance of one node) we cannot use the second order upwind scheme because $\nabla\tau$ changes signs. Therefore, for those nodes we simply use \eqref{eq:londDiff}. This situation is similar to the treatment of the same points when calculating $\tau$ using Fast Marching. There, the algorithm reverts to first order operators for these points, but still achieves second order accuracy overall; see \cite{TH2016}. For discretizing the advection term in \eqref{eq:ADR2} we use similar operators.

%

\subsubsection{Boundary conditions}
The boundary conditions of the ADR equation \eqref{eq:ADR} are identical to those of \eqref{eq:Helmholtz}, and are defined using the chain rule.
For example, for Neumann BC we have
\begin{equation}\label{eq:BCADR}
\bfn\cdot \nabla u = 0 \Rightarrow \bfn\cdot(\nabla a - i\omega \nabla\tau)\exp(-i\omega\tau) = 0 \Rightarrow \bfn\cdot\nabla a - i\omega\bfn\cdot\nabla\tau=0.
\end{equation}
The Sommerfeld BC in Eq. \eqref{eq:sommerfeld} is treated in a similar way, and the absorbing boundary layer in $\gamma$ is kept as a mass matrix similarly to the case of \eqref{eq:Helmholtz}. The boundary condition in Eq. \eqref{eq:BCADR} or its Sommerfeld version are needed for the Laplacian operator in \eqref{eq:ADR}-\eqref{eq:ADR2}, and can also be used for the advection term in this equation. Alternatively, the advection term can be discretized without boundary conditions. Since the source is inside the domain, the wave propagate from the source outwards, and the upwind discretization at the boundary points will always use internal neighboring points in \eqref{eq:londUpwindDiff}. Therefore, the upwind advection term in \eqref{eq:londUpwindDiff} does not require boundary conditions. This principle is used in Fast Marching to calculate $\tau$. On the other hand, the scheme \eqref{eq:londDiff} does require a different treatment at the boundaries and may be replaced with the upwind operators there (which is what we do here). Other boundary conditions for \eqref{eq:Helmholtz} can be adapted to \eqref{eq:ADR} similarly to \eqref{eq:BCADR}.

\subsection{The relation between the discretized versions of the ADR and Helmholtz equations}

The equations \eqref{eq:Helmholtz} and \eqref{eq:ADR} are diagonally scaled versions of each other \cite{haber2011fast}. For the matrix in \eqref{eq:HelmholtzLinSystem} and both matrices $\hat H$ in \eqref{eq:ADRLinSystemCentral}-\eqref{eq:ADRLinSystemUpwind}, we have that
\begin{equation}\label{eq:matrix_rescaling}
M^{-1}HM\approx \hat H,
\end{equation}
where $M$ is a diagonal matrix such that $M_{jj}=\exp(-i\omega\tau(x_j))$. We now show a comparison between the discrete matrices $M^{-1}HM$ and $\hat H_{cen}$ in 1D, and again the extensions to 2D and 3D are straightforward.

Assume that we know $\tau(x)$ for which \eqref{eq:eikonal} is held. A general equation of the system $M^{-1}HM\bfa = \hat\bfq$ is given by
\begin{equation}\label{eq:helmADR2}
\frac{1}{h^2}\left(a_{j-1}\exp(i\omega(\tau_j-\tau_{j-1}))-2a_{j}+a_{j+1}\exp(i\omega(\tau_j-\tau_{j+1}))\right)+\omega^2\kappa_j^2a_j = \hat q_j
\end{equation}
Using the Taylor expansion, we set : $\tau_{j\pm1} - \tau_j \approx \pm\tau'_jh + \frac{1}{2}\tau''_jh^2 + O(h^3)$ and get
\begin{eqnarray}\label{eq:helmADR3}
  \nonumber  \frac{1}{h^2}\left(a_{j-1}\exp(i\omega(-\tau'_jh + \frac{1}{2}\tau''_jh^2 + O(h^3)))-2a_{j}\right) + \quad&&\\
  \frac{1}{h^2}\left(a_{j+1}\exp(i\omega(\tau'_jh + \frac{1}{2}\tau''_jh^2 + O(h^3)))\right) +\omega^2\kappa_j^2a_j &=&\hat q_j.
\end{eqnarray}
\noindent Next, we use the Taylor expansion $\exp(\varepsilon)= 1 + \varepsilon + \frac{1}{2}\varepsilon^2 + \frac{1}{6}\varepsilon^3 + O(\varepsilon^4)$, for which we take many terms because the expression inside the exponent is of magnitude $\omega h$, which is typically small but much larger than $h$. We get
\begin{equation}\label{eq:helmADR4}
\begin{array}{ll}
\frac{1}{h^2}\left(a_{j-1} -2a_{j} +a_{j+1} \right)\\
 -\frac{i\omega}{h^2}\left((h + \frac{1}{6}\omega^2h^3)\tau'_j(a_{j+1}-a_{j-1})  + h^2\tau''_j\frac{1}{2}(a_{j+1}+a_{j-1})\right) \\
  - \frac{\omega^2}{h^2}\left( \frac{1}{2}(-\tau'_jh)^2a_{j-1}  + \frac{1}{2}(\tau'_jh )^2a_{j+1} - \omega^2h^2\kappa_j^2a_j\right) = \hat q_j,
\end{array}
\end{equation}
while neglecting terms which are of magnitude lower than $O(\omega^2 h^2)$ relatively to the magnitude of the entries of $H$.

The first and second lines of \eqref{eq:helmADR4} are identical to those in our discretization of \eqref{eq:ADR} using \eqref{eq:londDiff}. The second line suggests that the mass matrix that multiplies $\Delta\tau$ should be discretized with an averaging operator $\frac{1}{2}(a_{j+1} + a_{j-1})$ if we wish to make the two discretizations more similar.
Looking into third row of \eqref{eq:helmADR4}, neglecting everything except the leading $O(h^2)$ terms, we get $(\tau'_j)^2\frac{1}{2}(a_{j-1} + a_{j+1}) - \kappa_j^2a_j$, which leads to the eikonal equation. If we assume that eikonal equation $(\tau'_j)^2 = \kappa_j^2$ is held, we can replace this term with $\frac{1}{2}\omega^2\kappa_j^2(a_{j-1} - 2a_j+ a_{j+1})$, which is the discrete Laplacian of $a$ multiplied by some factor. To conclude, if we neglect the third term of the Taylor expansion for $exp()$, then the discretization of \eqref{eq:ADR} that is close to $M^{-1}HM$ given an exact $\tau$ is
\begin{equation}
\begin{array}{ll}
\frac{1}{h^2}\left(a_{j-1} -2a_{j} +a_{j+1} \right)(1-\frac{1}{2}h^2\omega^2\kappa_j^2)\\
 -i\omega\left(2\tau'_j(\frac{a_{j+1}-a_{j-1}}{2h})  + \tau''_j\frac{1}{2}(a_{j+1}+a_{j-1})\right) = \hat q_j.
\end{array}
\end{equation}
In light of this comparison, we use an averaging mass matrix in the mass term $i\omega(\Delta\tau)a$.

The comparison above shows that there is an $O(\omega^2h^2)$ difference between the two discretizations, which means that the solution of the two systems will be similar only if $\omega h$ is small enough. This is also a requirement that we have for discretizing \eqref{eq:Helmholtz} using standard methods. In fact, \cite{bayliss1985accuracy} states that the discretization error in the Helmholtz computations is of size $O(\omega^3h^2)$,  which imposes an even stronger requirement on the mesh size than a small $\omega h$.

\section{Multigrid solvers for the advection-diffusion-reaction equations}
\label{sec:MG}
In this section we describe the multigrid approaches that we use for solving \eqref{eq:ADRLinSystemCentral}-\eqref{eq:ADRLinSystemUpwind}, and for solving \eqref{eq:HelmholtzLinSystem} using the shifted Laplacian method. Generally, multigrid approaches aim at solving linear systems
$$
A\bfx=\bfb
$$
iteratively by using two complementary processes: relaxation and coarse grid correction. The relaxation is obtained by a standard iterative method like Jacobi or Gauss-Seidel, which is only effective for reducing error that is spanned by the eigenvectors of $A$ that correspond to relatively high eigenvalues (in magnitude). The remaining error, called ``algebraically smooth'', is spanned by the eigenvectors of $A$ corresponding to small eigenvalues (in magnitude), i.e., vectors $\bfe$ s.t.
\begin{equation}\label{eq:algSmooth}
\|A\bfe\| \ll \|A\|\|\bfe\|.
\end{equation}
To reduce this algebraically smooth error, multigrid methods use a coarse grid correction, where the error $\bfe$ for some iterate $\bfx^{(k)}$ is estimated by solving a coarser system
$$A_c\bfe_c = \bfr_c = P^\top(\bfb-A\bfx^{(k)}).$$
The matrix $A_c$ is an approximation of the matrix $A$ on a coarser grid (the subscript $c$ denotes coarse components). The matrix $P$ is a transfer operator that is used for projecting the residual onto the coarser grid, and interpolating $\bfe_c$---the solution of the coarse system---back onto the fine grid, that is:
\begin{equation}
\bfe = P\bfe_c.
\end{equation}
This process is effective if any algebraically smooth error $\bfe$ satisfying \eqref{eq:algSmooth} can be represented in the range of the interpolation $P$.
The coarse operator $A_c$ can be obtained by either rediscretizing the problem on a coarser grid or by the Galerkin operator
\begin{equation}\label{eq:coarseGridMatrix}
A_{c} = P^\top AP.
\end{equation}
Algorithm \ref{alg:TwoCycle} summarizes the process using two grids. By treating the coarse problem recursively, we obtain the multigrid V-cycle, and by treating the coarse problem recursively twice (by two recursive calls to V-cycle) we obtain a W-cycle. For more information see \cite{BHM00,TOS01,Yav06} and references therein.

\begin{algorithm}
\DontPrintSemicolon
Algorithm $\bfx\leftarrow TwoGrid(A,\bfb,\bfx).$
\begin{enumerate}
\item Apply pre-relaxations: $\bfx \leftarrow Relax(A,\bfx,\bfb)$\;
\item Define and restrict the residual $\bfr_c = P^T(\bfb - A\bfx)$.
\item Define $\bfe_c$ as the solution of the coarse-grid problem $A_c\bfe_c=\bfr_c$.
\item Prolong $\bfe_c$ and apply coarse grid correction: $\bfx \leftarrow \bfx + P\bfe_c$.
\item Apply post-relaxations: $\bfu \leftarrow Relax(A,\bfx,\bfb)$.
\end{enumerate}
\caption{Two-grid cycle.}
\label{alg:TwoCycle}
\end{algorithm}

\subsection{The Shifted Laplacian multigrid method}
The shifted Laplacian multigrid method is one of the most common approaches to solve the Helmholtz equation. This method is implemented in efficient software packages \cite{knibbe2011gpu,Tobias2017}. To solve the linear system \eqref{eq:HelmholtzLinSystem} using the shifted Laplacian approach, one introduces a shifted system by adding a complex negative mass matrix to
\eqref{eq:HelmholtzLinSystem}
\begin{equation}\label{eq:shift}
H_s = H - i\omega^2\alpha\mbox{diag}(\kappa^2),
\end{equation}
where $\alpha>0$ is a shifting parameter. From a physical point of view this term is equivalent to adding attenuation to the equation, which means that the waves in the shifted problem decay rapidly if $\alpha$ is large enough, resulting in a \emph{local} approximation of the waveform. In the shifted Laplacian approach, the shifted matrix is used as a preconditioner for the Helmholtz linear system \eqref{eq:HelmholtzLinSystem} inside a Krylov method, which is usually chosen to be (flexible) GMRES \cite{saad1993flexible}. The preconditioning is obtained by applying a multigrid cycle for approximately inverting the shifted matrix \eqref{eq:shift}. The larger $\alpha$ the more efficient is the solution of the \emph{shifted} system using multigrid, but the quality of the \eqref{eq:shift} as preconditioner deteriorates. The compromise suggested in \cite{erlangga2006novel} is to use $\alpha = 0.5$, together with rather modest $F(1,1)$ cycles, while \cite{calandra2013improved} and \cite{JointEikFWI17} suggest applying more elaborate cycles.

We define the prolongation $P$ to be a bilinear interpolation operator, because the Laplacian operator in \eqref{eq:Helmholtz} is homogenous (does not have varying coefficients).
As relaxation, the damped Jacobi method is often chosen, and its damping parameter needs to be chosen differently for each level, as on coarse grids the wave number becomes larger compared to the mesh size. Consequently, the matrix becomes more indefinite, and even negative definite at some level \cite{elman2001multigrid,calandra2013improved}. Another choice of relaxation is the GMRES method, which automatically adapts to the matrix at each level. This relaxation method was originally suggested in \cite{elman2001multigrid} for the Helmholtz equation, and was recently used in \cite{cools2014new} and \cite{calandra2013improved}.

Beside the type of relaxation and prolongation, one has to choose the number of levels used in the multigrid hierarchy. Unlike many other multigrid scenarios, the algebraically smooth error modes of the Helmholtz operator have a sign-changing behavior at high wave number, and therefore cannot be represented well on very coarse grids. Hence, the performance of the solver deteriorates when using more levels. For example, the results in \cite{calandra2013improved} show that the best performance is achieved using three levels only, which is also what we get in our experience. The works in \cite{calandra2013improved} and \cite{erlangga2008multilevel} invest more work on the second grid, as oscillatory errors are significantly better represented on this grid than on the other coarser grids. However, when using only a few levels we get a rather large coarsest grid problem, which requires a relatively accurate solution. Factorizing the coarsest grid matrices for large scale 3D problems is memory consuming and limiting. The work in \cite{calandra2013improved} suggests an inexact solution of the coarsest grid using GMRES, which is the approach that we adopt in this work.

Another acceleration technique, suggested in \cite{erlangga2008multilevel}, uses a recursive multilevel Krylov solver for the coarse grid problems. Such a cycle is called Krylov-cycle, and has a rather elaborated recursive structure, depending on the number of Krylov iterations at each levels. If this number is 2 (which is the common choice), we get the structure of a W-cycle---see Algorithm \ref{alg:KrylovCycle}.

\begin{algorithm}
\DontPrintSemicolon
Algorithm $\bfx\leftarrow Kcycle(A,\bfb,\bfx)$
\begin{enumerate}
\item Apply pre-relaxations: $\bfx \leftarrow Relax(A,\bfx,\bfb)$\;
\item Define and restrict the residual $\bfr_c = P^T(\bfb - A\bfx)$.
\item If coarsest level is reached - solve $A_c\bfe_c= \bfr_c$, possibly inexactly.\\
Otherwise, apply FGMRES(2) for $A_c\bfe_c= \bfr_c$ starting from 0,\\ with $Kcycle()$ as a preconditioner.
\item Prolong $\bfe_c$ and apply coarse grid correction: $\bfx \leftarrow \bfx + P\bfe_c$.
\item Apply post-relaxations: $\bfu \leftarrow Relax(A,\bfx,\bfb)$.
\end{enumerate}
\caption{Krylov multigrid cycle}
\label{alg:KrylovCycle}
\end{algorithm}

\subsection{The solution of the ADR linear system}
\begin{figure}
\centering
\includegraphics[width=0.9\textwidth]{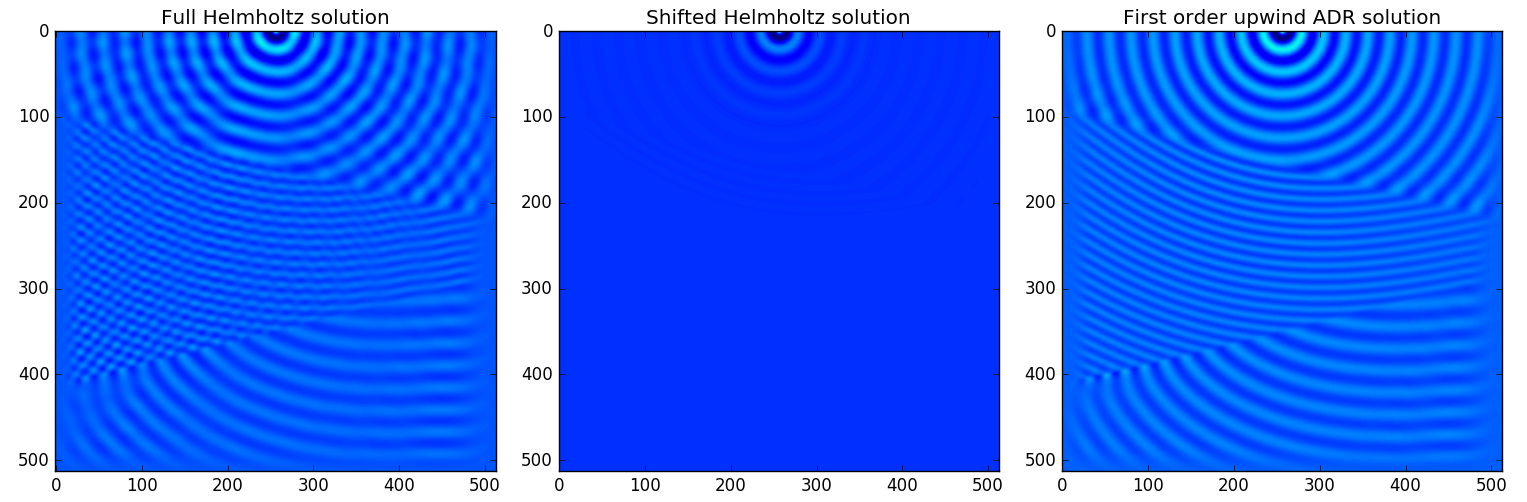}
\caption{A demonstration of the local and global approximations of the Helmholtz solution. On the left there is a solution to \eqref{eq:Helmholtz} for point source with a reflective model (the Wedge model which appears later). Indeed, reflections are evident in the middle and upper parts of the domain. In the middle figure we show the solution obtained by solving the shifted problem \eqref{eq:shift} for the same point source, for $\alpha=0.2$. It is clear that the waves decay rapidly, and only 2-3 wavelengths are approximated well. This is the reason why the Shifted Laplacian method is so sensitive to the number of wavelengths in the domain. On the left we see the solution of \eqref{eq:1up} for the point source, which can be obtained almost as easily as the solution of \eqref{eq:shift}. This time we see the global main wave function on the entire domain, but the discretization eliminates all the reflections.   }
\label{fig:localGlobal}
\end{figure}

We have two ADR linear systems in \eqref{eq:ADRLinSystemCentral}-\eqref{eq:ADRLinSystemUpwind}. The work of \cite{haber2011fast} suggests to use $M\hat HM^{-1}$ as a preconditioner to \eqref{eq:HelmholtzLinSystem} where $\hat H$ is one of the ADR matrices. From the results and local Fourier analysis in \cite{haber2011fast} we learn the following:
\begin{enumerate}
\item Even at high frequency, the operator $M\hat H_{cen}M^{-1}$ is a good preconditioner to $H$, and the two matrices are spectrally similar. The linear system \eqref{eq:ADRLinSystemCentral} is hard to solve using multigrid, similarly to the standard \eqref{eq:HelmholtzLinSystem}.
\item The ADR operator $M\hat H_{2up}M^{-1}$ is not a good preconditioner to $H$, suggesting that the operators are indeed different. However, the ADR linear system \eqref{eq:ADRLinSystemUpwind} can be efficiently solved using multigrid.
\end{enumerate}
Even though we use a significantly different $\tau$ than \cite{haber2011fast}, we observed the same properties in our case as well. In addition, the ADR system with a \emph{first} order upwind advection operator is solved very efficiently by multigrid. This is not surprising because the first order advection operator can be obtained as a sum of a second order advection operator and a Laplacian operator which is well represented on coarser grids (for both upwind and central schemes). This is similar to having a ``Laplacian shift'' term of $-i\omega h\Delta a$. Unlike the standard shift in \eqref{eq:shift}, the ``Laplacian shift'' strongly damps oscillatory modes in $a$, but hardly influences spatially smooth modes.
Because we solve the Helmholtz equation for a point source and use an accurate travel time, most of the amplitude $a$ is smooth (up to reflections which are oscillatory). This results in a \emph{global} approximation which has the global behavior of the solution corresponding to the first arrival of the wave, but has almost no reflections. In Fig. \ref{fig:localGlobal} we demonstrate such a global approximation compared with a local approximation obtained by the shifted Laplacian preconditioner.

We solve the system \eqref{eq:ADRLinSystemCentral} in two stages. In the first stage we compute the global approximation of the solution which corresponds to the first arrival of the wave without treating the reflections well. This is efficiently obtained by approximately solving
\begin{equation}\label{eq:1up}
\hat H_{1up}\bfa = \hat\bfq,
\end{equation}
up to a quite low accuracy. Once \eqref{eq:1up} is approximately solved, we finalize the solution and add the missing reflections. To this end, we use the approximate solution of \eqref{eq:1up} as an initial guess, and apply the shifted Laplacian approach to the ADR system rescaled as Helmholtz
\begin{equation}\label{eq:ADRcenRescaled}
M\hat H_{cen} M^{-1} \bfu = \bfq,
\end{equation}
which is the opposite of $\eqref{eq:matrix_rescaling}$. Solving \eqref{eq:1up} inaccurately is achieved by Algorithm \ref{alg:KrylovCycle} in very few iterations, so the most of the work in obtaining the solution is invested in the second stage. We note that in principle, we can use only the shifted Laplacian approach to solve \eqref{eq:ADRLinSystemCentral} without the first stage, and the opposite is also true---we can use the two stages approach to treat \eqref{eq:HelmholtzLinSystem} using the opposite rescaling.

\begin{algorithm}
\DontPrintSemicolon
Preprocessing: Solve the factored eikonal equation and obtain the travel time $\tau$.\;
 \emph{\# Stage 1: define a global approximate solution}
 \begin{enumerate}
    \item Compute $\bfa_{1up}$ as a low-accuracy solution of $\hat H_{1up}\bfa = \hat\bfq$.\;
    \item Define $M = \mbox{diag}(\exp(-i\omega\tau))$, $\bfu_{1up} = M \bfa_{1up}$.
 \end{enumerate}
\emph{\# Stage 2: Complete the solution using the shifted Laplacian method}\vspace{-7pt}
 \begin{enumerate}
 \item Define the shifted operator: $H_s^{cen} = M\hat H_{cen} M^{-1}- i\omega^2\alpha\mbox{diag}(\kappa^2)$.\;
 \item Solve $M\hat H_{cen} M^{-1}\bfu = \bfq$ using $H_s^{cen}$ as preconditioner with multigrid, starting from $\bfu_{1up}$.
\end{enumerate}
\caption{The two-stage solution of the central difference ADR equation.}
\label{alg:ADRlong}
\end{algorithm}

To solve the ADR system \eqref{eq:ADRLinSystemUpwind} using second order upwind discretization we use a preconditioner matrix
\begin{equation}\label{eq:precADRup}
(1-\beta)\hat H_{2up} + \beta\hat H_{1up}
\end{equation}
where $\hat H_{1up}$ is the ADR system with first order upwind advection in Eq. \eqref{eq:1up}. We treat this preconditioner using Algorithm \ref{alg:KrylovCycle}.

\section{Numerical results}\label{sec:results}
In this section we compare two aspects of the ADR approach for solving the Helmholtz problem. First, we compare accuracy of the numerical solution $u(\vec{x})$ of the Helmholtz equation \eqref{eq:Helmholtz} when it is discretized using the standard and ADR second order discretizations, where $u(\vec{x})$ is composed of the solutions of \eqref{eq:ADR} and \eqref{eq:eikonal} through \eqref{eq:new_u}. To obtain the travel time $\tau$ we use the Fast Marching algorithm in \cite{TH2016} using a second order upwind discretization. Second, we compare the computational effort required to obtain the different solutions using multigrid methods. 

\subsection{Accuracy comparison}
\begin{figure}
\centering
\includegraphics[width=1.0\textwidth]{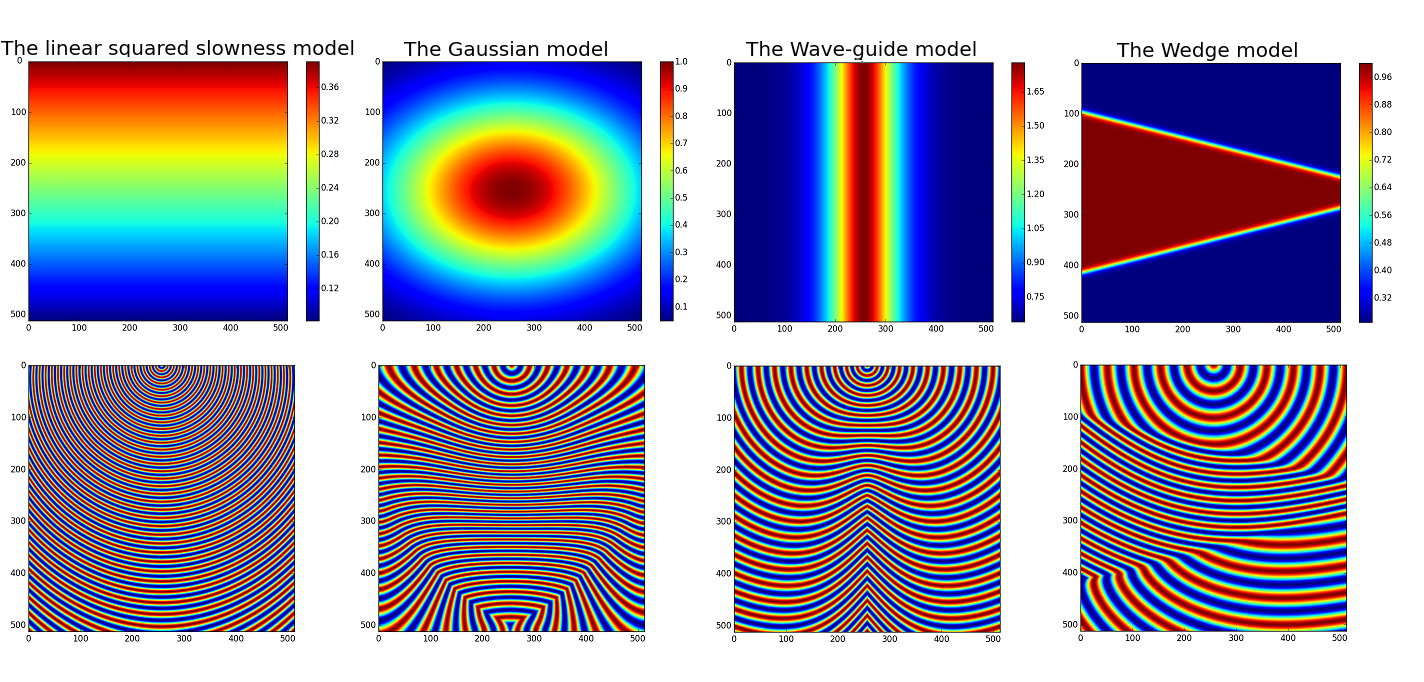}
\caption{The four test cases for the accuracy comparison. In the upper row we show the model $\kappa(\vec x)^2$ for each test. From left to right: the linear squared slowness model, the Gaussian Model, the wave-guide model and the wedge model. On the bottom row we show the phase according to the travel time $\tau$ corresponding for each model.}
\label{fig:AccuracyTests}
\end{figure}

In this section we empirically compare the accuracy of the different discretizations described in Section \ref{sec:discretization}. We test the accuracy by solving a given problem with all discretizations, and comparing the solutions to a reference solution obtained on a four times finer grid using the standard second order discretization \eqref{eq:HelmholtzLinSystem}. Some of the test cases that we present involve caustics and reflections.

We consider four heterogenous test cases, all on a 2D unit square, discretized on a nodal regular $513\times513$ grid. Motivated by geophysical applications, we place the source point on the top row of the model (the surface), where we use Neumann boundary conditions. On the bottom and sides of the model we use an absorbing boundary condition to prevent reflection from the model boundaries, as the waves are supposed to continue spreading from these boundaries. We compare the obtained solutions to a reference solution $\bfu_{ref}$ calculated for the same problem on a $2049\times2049$ grid, and then downsampled. The $\delta$ function at the source is discretized as $\frac{1}{h^2}$.  We consider the four models shown in top row of Fig. \ref{fig:AccuracyTests}, where on the bottom are the corresponding phases $\exp(-i\omega\tau)$. For each discretization---the standard $2^{nd}$ order finite difference, the ADR with cental difference advection, and the ADR with $2^{nd}$ order upwind advection---we show the obtained solution $\bfu$ and the \emph{relative} error
$$
e_{ij} = \frac{|u_{ij} - (u_{ref})_{ij}|}{|(u_{ref})_{ij}|}.
$$

\subsubsection{Linear squared slowness model}
\begin{figure}
\centering
\includegraphics[width=1.0\textwidth]{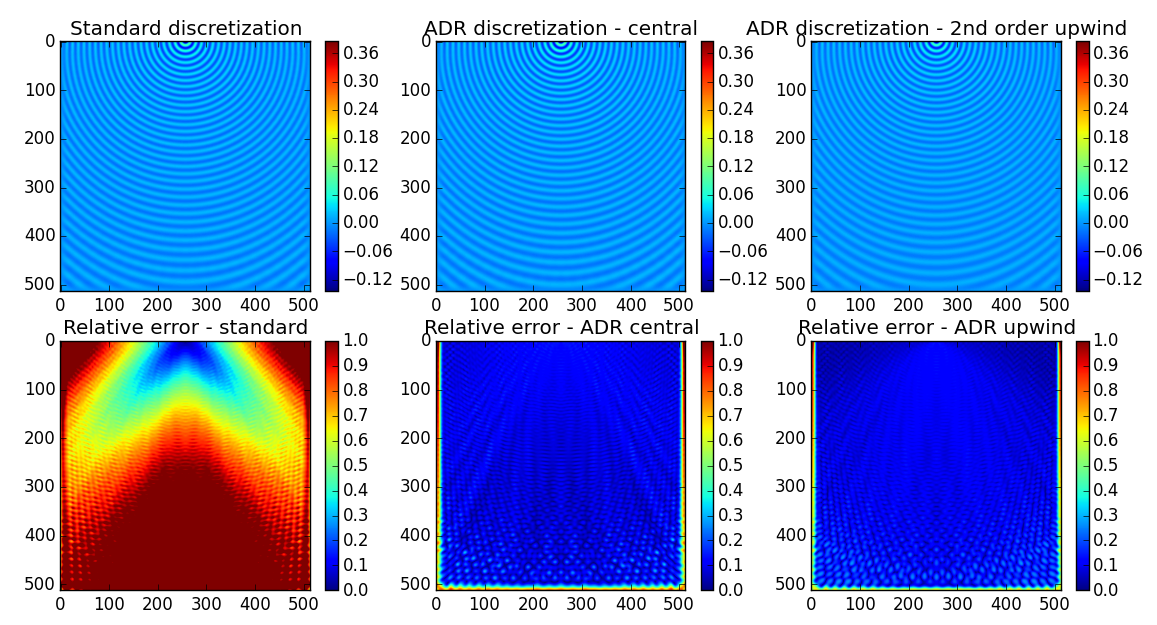}
\caption{Accuracy comparison for the linear model. Note that the error that introduced by the standard discretization is mostly a phase error, which is not observed in the other ADR discretizations.}
\label{fig:LinearAccuracy}
\end{figure}
In this test case the squared slowness $\kappa(\vec x)^2$ is a linear model in the $y$ direction, starting from 0.4 at the top of the model and decreases to 0.08 at the bottom. This model is the first model from the left in Fig. \ref{fig:AccuracyTests}. For this test case we use a high value of $\omega$ corresponding to at least 11 grid points per wavelength. Fig. \ref{fig:LinearAccuracy} shows the real value of $\bfu$ for the three discretizations, and the relative absolute error. The solution for this test does not contain reflections or caustics, and therefore the amplitude and travel time are smooth. Visually, the three solutions are similar, but the error plot shows a very high error for the standard discretization. This is a phase error that is caused by the dispersion phenomenon mentioned earlier. The ADR discretizations do not include the dispersion because the phase is obtained accurately by the relatively smooth $\tau$. This example illustrates that when there are no reflections or caustics, and the travel time is smooth (except at the point source), the solution for the amplitude obtained by the ADR equation is more accurate than the standard approach.

\subsubsection{Gaussian model}

In this test case the squared slowness $\kappa(\vec x)^2$ is a Gaussian function
$$
\kappa(\vec{x})^2 = \exp(-(\vec{x}-0.5)^\top\Sigma(\vec{x}-0.5)), \quad \Sigma = \left[\begin{array}{cc}4 & 0 \\ 0 & 8\end{array}\right],
$$
\noindent The model appears second from the left in Fig. \ref{fig:AccuracyTests}, and the corresponding travel time $\tau$ has a discontinuity at the bottom of the model. For this test case we choose $\omega$ so we have at least 12 grid points per wavelength.

\begin{figure}
\centering
\includegraphics[width=1.0\textwidth]{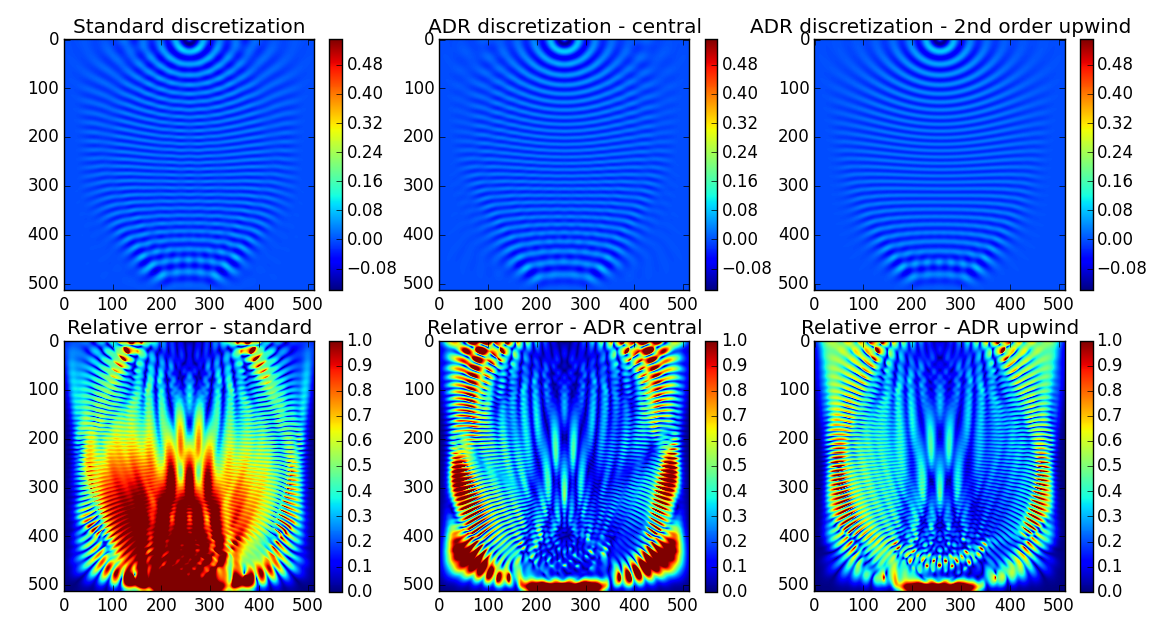}
\caption{Accuracy comparison for the Gaussian model.}
\label{fig:GaussianAccuracy}
\end{figure}

Fig. \ref{fig:GaussianAccuracy} shows the obtained solutions, which are visually similar. Because of the caustics, this time all approximations have significant errors, but the error is more dominant when using the standard discretization than in the ADR discretizations. The upwind ADR discretization yields the most accurate solution in general, but is comparable to the central difference ADR solution. The standard discretization again yields a solution with dispersion errors because the frequency is high.

\subsubsection{The wave-guide model}
In this test case the squared slowness $\kappa(\vec x)^2$ is the waveguide model given by
$$
v(\vec x) = \exp(1.25*(1-0.4*\exp(-32*(x_1 - 0.5)^2)));\quad \kappa(\vec x)^2 = \frac{1}{v^2}.
$$
The model is shown second from the right in Fig. \ref{fig:AccuracyTests}. Even though the model is very smooth, it generates severe caustics, which leads to inferences in the solution. Since this test case is very complicated, we use a modest $\omega$ with at least 20 grid points per wavelength.
\begin{figure}
\centering
\includegraphics[width=1.0\textwidth]{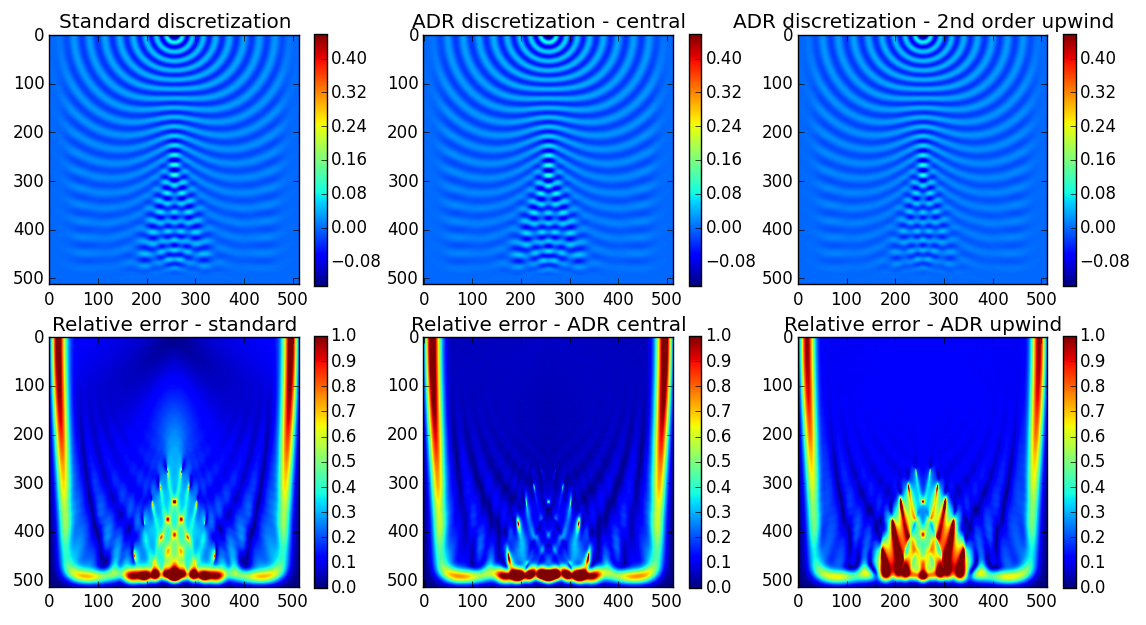}
\caption{Accuracy comparison for the wave-guide model.}
\label{fig:WaveGuideAccuracy}
\end{figure}

Fig. \ref{fig:WaveGuideAccuracy} shows the obtained solutions, and in particular it shows the inferences at the bottom part. All solutions are again visually similar. Because of the caustics there are errors, and this time it is clear that the ADR discretization with central difference advection has the lowest error of the three, and the ADR with upwind advection has the highest error compared with the reference solution. Because the frequency is not so high, the standard discretization does not introduce dispersion errors.

\subsubsection{The wedge model}

This model, which is the rightmost model in Fig. \ref{fig:AccuracyTests}, is usually given with a sharp step. Here, we smooth the step to be able to have a somewhat accurate fine approximation of the solution as a reference. The top half of the model is given by the function
$$
\kappa^2 = 0.25*(\tanh((4*x_2-x_1 - 0.75)*20)) + 0.75,
$$
and the bottom half is generated by mirroring the top half. In our experiment, the step is smoothed over approximately 20 grid-points, for the coarser $513\times513$ grid. The frequency chosen for the experiment is $f = 20 Hz$ ($\omega = 2\pi f$), and the wavelength is at least 25 grid points. This model generates reflections from the wedge and is hard to model accurately because of the sharp change in the model.

\begin{figure}
\centering
\includegraphics[width=1.0\textwidth]{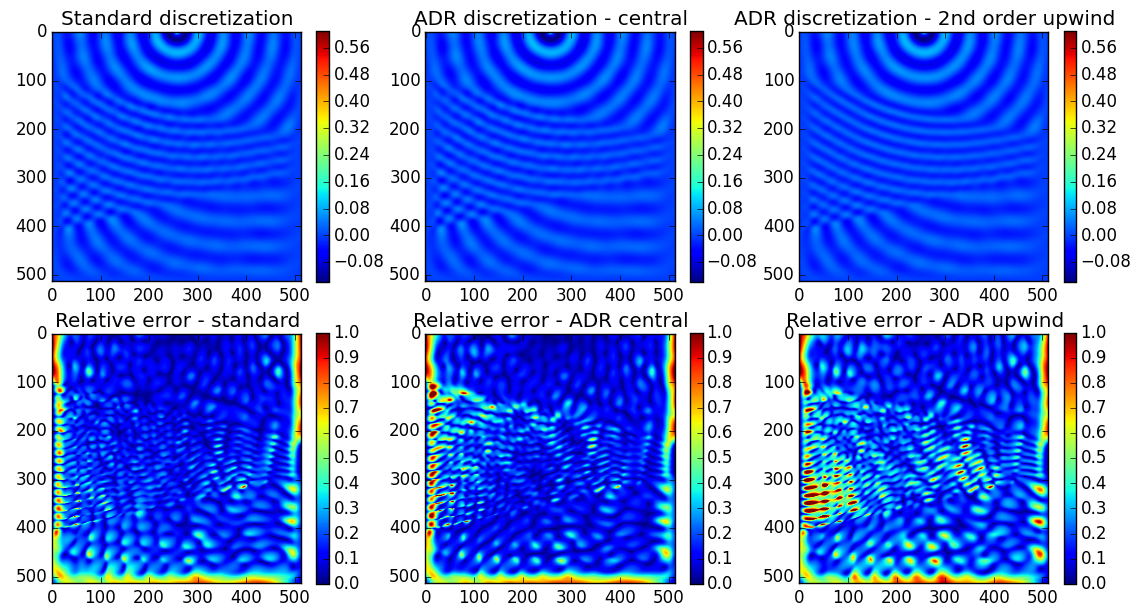}
\caption{Accuracy comparison for the wedge model.}
\label{fig:WedgeAccuracy}
\end{figure}

Fig. \ref{fig:WedgeAccuracy} shows the obtained solutions, and in particular it shows the reflections at the top and middle parts of the model. Visually, the two left approximations are similar, but the right one includes somewhat weaker reflections at the middle part of the model. Besides that, all solutions show comparable errors with the lowest error obtained by the standard discretization. We note that particularly in this test case, we are not certain that reference solution accurately models the reflections.

\subsection{Numerical solution performance}

In this section we compare the computational effort required to solve the Helmholtz problem with the 3 discretizations described earlier. We present examples that appear in geophysical applications, where typically the length of the domain is long, and the depth of the domain is rather short. We consider two and three dimensional examples, for which we test the performance of the solver at high frequency (10-12 grid points per wavelength), and intermediate frequency (about 15-17 grid points per wavelength). For each case we present three grid sizes to demonstrate the scalability of the solvers as the mesh size and frequency grow together. In all cases, we use flexible GMRES(5) with multigrid as a preconditioner and seek a solution with relative residual accuracy of $10^{-5}$, starting from a zero initial guess. For solving the linear system \eqref{eq:HelmholtzLinSystem} arising from the standard discretization we use the shifted Laplacian preconditioner with a shift $\alpha=0.2$. We use a geometric multigrid configuration that includes the Krylov multigrid cycles in Algorithm \eqref{alg:KrylovCycle}, with Jacobi-preconditioned GMRES as relaxation. We define our coarse grid problems by the Galerkin product $P^\top H P$. As the problem becomes more indefinite on coarser grids, we found that it is worthy applying more relaxations on the coarser grids, and therefore on level $l$ we apply $l+1$ pre- and post-relaxations. That is, 2 for the first level, 3 for the second and so on. Having large 3D problems in mind, we use 5 levels and approximate coarsest grid solution. As in \cite{calandra2013improved}, we apply 10 Jacobi-preconditioned GMRES iterations instead of a direct solve using a factorization. Because we use relatively elaborate and expensive cycles, we are able to solve the Helmholtz problem using much less cycles compared to using more standard cycles with $\alpha=0.5$.

We use the same multigrid cycles for the ADR discretizations. For solving \eqref{eq:ADRLinSystemUpwind} we use \eqref{eq:precADRup} as a preconditioner treated by multigrid, and choose $\beta = 0.25$. This seems to be the most effective option together with our elaborated cycles for test cases at rather large scales. For solving \eqref{eq:ADRLinSystemCentral} we apply the two stage scheme described in Algorithm \ref{alg:ADRlong}. For the first stage, we apply at most 5 Krylov-cycles, or stop earlier if the relative residual drop with respect to \eqref{eq:1up} is $10^{-2}$. Once the intermediate residual drop is reached, we switch to solving \eqref{eq:ADRcenRescaled} using the same shifted Laplacian configuration mentioned before until convergence, to capture the reflections in the solution.

We show the total number of Krylov-cycles required for convergence (\#$it$), the solution time ($t_{sol}$) and the time required to compute the travel time $\tau$ by Fast Marching ($t_{FM}$). Our code is written in Julia language \cite{Julia}, and is available as part of the jInv software \cite{jInv17} (see {\tt https://github.com/JuliaInv/ForwardHelmholtz.jl}). Our two dimensional tests were computed on a laptop machine using Windows 10 64bit OS, with Intel core-i7 2.8 GHz CPU with 32 GB of RAM. The three dimensional tests were computed on a workstation with Intel Xeon E5-2620 \@ 2GHz X 2 (6 cores per socket, 2 Threads per core for a total of 24 cores) with 64 GB RAM, running on Centos 7 Linux distribution. For the 3D results we use single precision computations, to save memory. The code for the solution phase is parallelized in the matrix-vector products, while the FM code is sequential as the algorithm is sequential. The parallelism can be efficiently exploited for FM if many linear systems need to be solved, and the travel times are calculated simultaneously \cite{jInv17}.

\subsubsection{2D linear model}

\begin{figure}
\centering
\includegraphics[width=0.9\textwidth]{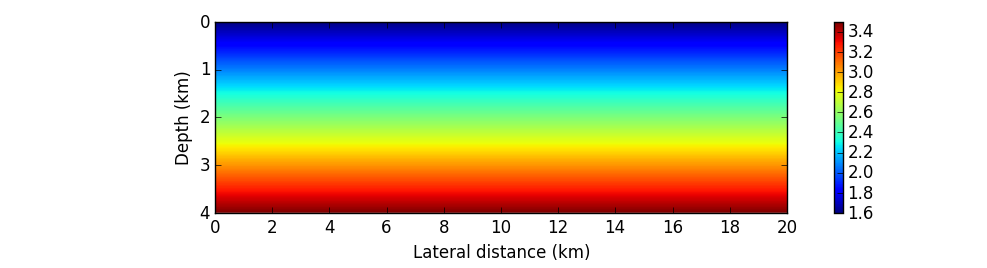}
\caption{The linear velocity model. Units are in $km/sec$, and $\kappa^2 = \frac{1}{v^2}$.}
\label{fig:LinearVel}
\end{figure}

Our first test case is a linear velocity model ($\kappa^2$ is the inverse squared velocity) that does not include reflections. The model is given in Figure \ref{fig:LinearVel}, and the results are summarized in Table \ref{tab:MGlinear}. It is clear that the shifted Laplacian method requires the most iterations to solve the problem. Moreover, it requires more iterations as the problem gets larger and the frequency is higher. The solution of \eqref{eq:ADRLinSystemCentral}, denoted as ``ADR central'', requires about 20\%-30\% less iterations and time, thanks to the global approximation obtained in the first phase. Solving \eqref{eq:ADRLinSystemUpwind} is achieved with the least iterations and time, and more importantly - it is fairly mesh and frequency independent. Since this model does not contain reflections or caustics, then the ``ADR upwind'' option is the best one because it provides both accurate approximation and fast solution. It is expected to continue being so at even larger scales.

\begin{table}
\centering
\small
\begin{tabular}{|c|c|c|cc|cc|cc|c|}
  \hline
   & $f (Hz)$ &Points per& \mc{2}{|c|}{\emph{Standard}}& \mc{2}{c|}{\emph{ADR central}} & \mc{2}{c|}{\emph{ADR upwind}} & \mc{1}{c|}{\emph{eikonal}}\\
    Grid size &$=\omega/2\pi$ &wavelength&  \#$it$  &  $t_{sol}$  & \#$it$  &  $t_{sol}$  & \#$it$  &  $t_{sol}$  & $t_{FM}$\\
  \hline
  \hline
  \mr{2}{$769\times257$} & 3.5 & 17.5 & 43 & 5.9s & 34 & 4.2s & 39 & 5.8s & 0.33s \\
                         & 5.5 & 11.2 & 67 & 9.1s &  54 & 6.5s  & 33 &4.7s& 0.37s  \\
  \hline
  \mr{2}{$1025\times385$} & 5.5 & 14.9 & 74 & 17.9s & 56 & 13.0s & 41 & 10.7s & 0.75s \\
                          & 7.5 & 10.9 & 102 & 24.8s & 74 & 17.0s & 40 & 10.4s & 0.82s \\
  \hline
  \mr{2}{$1537\times513$} & 7.5 & 16.3 & 93 & 42.3s & 64 & 28.0s & 46 & 22.6s & 1.64s \\
                          & 11.0 & 11.2 & 151 & 68.2s & 93 & 42.9s & 46 & 22.5s & 1.65s \\
  \hline
 \end{tabular}
\caption{Linear model: solution performance. }
\label{tab:MGlinear}
\end{table}


\subsubsection{The 2D Marmousi 2 model}

\begin{figure}
\centering
\includegraphics[width=0.9\textwidth]{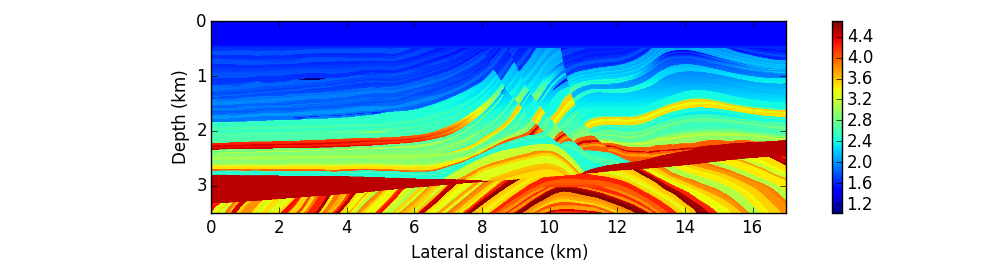}
\caption{The Marmousi2 P-wave velocity model. Units are in $km/sec$, and $\kappa^2 = \frac{1}{v^2}$.}
\label{fig:Marmousi2}
\end{figure}

Our second test case is the P-wave velocity of the Marmousi 2 model \cite{martin2006marmousi2}, which is given in Figure \eqref{fig:Marmousi2}. This model is mostly piecewise constant, and includes many reflectors. Table \ref{tab:MGmarmousi} summarizes the results for this test case. The performance of the shifted Laplacian approach is quite similar to the previous test case. The method is quite robust to the heterogeneity of the model and more sensitive to the frequency. In the ``ADR central'' section, we see less advantage compared to the previous smooth test case. That is because the ``global'' reflection-less approximation obtained from solving \eqref{eq:1up} is less effective. We note again that \eqref{eq:ADRLinSystemCentral} can be solved using the shifted Laplacian method alone with the same efficiency as \eqref{eq:HelmholtzLinSystem}. The third option ``ADR upwind'' is again achieved in less iterations and time, and is again more mesh independent than the other options. 

\begin{table}
\centering
\small
\begin{tabular}{|c|c|c|cc|cc|cc|c|}
  \hline
   & $f (Hz)$ &Points per&  \mc{2}{|c|}{\emph{Standard}}& \mc{2}{c|}{\emph{ADR central}} & \mc{2}{c|}{\emph{ADR upwind}} & \mc{1}{c|}{\emph{eikonal}}\\
    Grid size & $=\omega/2\pi$ & wavelength& \#$it$  &  $t_{sol}$  & \#$it$  &  $t_{sol}$  & \#$it$  &  $t_{sol}$  & $t_{FM}$\\
  \hline
  \hline
  \mr{2}{$769\times257$} &  3.0& 17.3& 42 & 5.7s & 35 & 4.6s & 41 & 6.1s & 0.46s \\
                         & 4.5 & 11.5& 64 & 8.8s &  50 & 7.0s  & 42 &6.2s & 0.47s  \\
  \hline
  \mr{2}{$1025\times385$} &  4.5 &15.4 & 61 & 16.5s & 52 & 13.1s & 55 & 14.6s & 0.93s \\
                          &   6.5 & 10.8& 115 & 28.1s & 93 & 22.4s & 52 & 13.9s & 0.84s \\
  \hline
  \mr{2}{$1537\times513$} & 6.5 & 16.0 & 95 & 42.8s & 79 & 36.1s & 63 & 31.4s & 1.8s \\
                          & 9.0 & 11.5 & 164 & 76.8s & 124 & 58.1s & 54 & 26.7s & 1.8s \\
  \hline
 \end{tabular}
\caption{Marmousi 2 model: solution performance. }
\label{tab:MGmarmousi}
\end{table}


\subsubsection{The 3D linear model}

\begin{table}
\centering
\small
\begin{tabular}{|c|c|c|cc|cc|cc|c|}
  \hline
   & $f (Hz)$ &Points per& \mc{2}{|c|}{\emph{Standard}}& \mc{2}{c|}{\emph{ADR central}} & \mc{2}{c|}{\emph{ADR upwind}} & \mc{1}{c|}{\emph{eikonal}}\\
    Grid size &$=\omega/2\pi$ &wavelength& \#$it$  &  $t_{sol}$  & \#$it$  &  $t_{sol}$  & \#$it$  &  $t_{sol}$  & $t_{FM}$\\
  \hline
  \hline
  \mr{2}{$257\times257\times65$}  & 1.5 & 13.6 & 18 & 115s &  18 & 87s & 17 & 107s & 16.5s \\
                                  & 2.0 & 10.2 & 22 & 140s &  23 & 110s  & 18 &116s& 17s  \\
  \hline
  \mr{2}{$385\times385\times97$}  & 2.0 & 15.3  & 23 & 212s & 21 & 160s & 19 & 170s & 60.5s \\
                                  & 3.0 & 10.2 & 33 & 302s & 32 & 242s & 20 & 180s & 62s \\
  \hline
  \mr{2}{$513\times513\times129$} & 3.0 & 13.7 & 35 & 534s & 30 & 401s & 21 & 311s & 165s \\
                                  & 4.0 & 10.2 & 41 & 620s & 36 & 490s & 24 & 352s & 159s \\
  \hline
 \end{tabular}
\caption{3D Linear model: solution performance.  }
\label{tab:Linear3D}
\end{table}

In this experiment we consider the three dimensional version of the linear model in Fig. \ref{fig:LinearVel}, which is smooth and does not introduce reflections. Table \ref{tab:Linear3D} summarizes the results for this test case. Because of memory limitations, the resolution and frequencies are much lower here than in 2D, and therefore, the iteration counts are much lower than in 2D. In terms of iteration counts, we again see that the standard discretization is taking the most iterations, while the other ADR systems are solved in less iterations (with upwind ADR taking the least). The time-per-iteration in the ADR central column are lower than in the standard column, probably because of less work that is done on coarser grids. That is, in Algorithm \ref{alg:KrylovCycle}, some cycles do not include a second recursive call because a threshold is achieved in FGMRES (we use 0.1 as a threshold for the coarse grid as suggested in \cite{notay2008recursive}).  In light of the 2D results, we do not expect this to be a significant advantage of the ADR discretization over the standard discretization in larger problems. The cost of the FM preprocessing is higher now compared to the 2D case, because in addition to the problem size, the algorithm is more complicated in this case. Still, the solution time of FM is not very significant, considering that the multigrid solution timings are obtained using a highly parallelized code (16 cores), while FM is completely serial. This can be exploited if many systems are required to be solved. As the problem gets larger, and the multigrid solution takes more and more iterations, the FM cost will become less and less significant. \newline \textit{Remark}: although the cost of each cycle in our solver is of linear complexity, the timings in Table \ref{tab:Linear3D} scale slightly better than linearly. That is a result of better performance in the parallelism of our code---as the problem grows, the efficiency of the matrix-vector multiplications is better. This behaviour is obviously independent of the method.

\subsubsection{The 3D Overthrust model}
\begin{figure}
\centering
\includegraphics[width=0.75\textwidth]{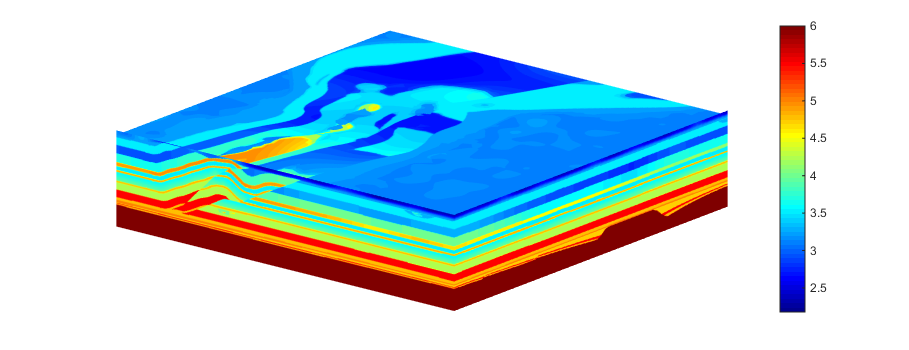}
\caption{The SEG Overthrust velocity model. Units are in $km/sec$, and $\kappa^2 = \frac{1}{v^2}$. The model corresponds to a domain of $20\times20\times4.65$ km.}
\label{fig:Overthrust}
\end{figure}

\begin{table}
\centering
\small
\begin{tabular}{|c|c|c|cc|cc|cc|c|}
  \hline
   & $f (Hz)$ &Points per& \mc{2}{|c|}{\emph{Standard}}& \mc{2}{c|}{\emph{ADR central}} & \mc{2}{c|}{\emph{ADR upwind}} & \mc{1}{c|}{\emph{eikonal}}\\
    Grid size &$=\omega/2\pi$ &wavelength& \#$it$  &  $t_{sol}$  & \#$it$  &  $t_{sol}$  & \#$it$  &  $t_{sol}$  & $t_{FM}$\\
  \hline
  \hline
  \mr{2}{$257\times257\times65$}  & 2.0 & 14.4 & 14 & 89s &  13 & 62s & 19 & 118s & 16.5s \\
                                  & 3.0 & 9.6  & 20 & 127s &  19 & 91s  & 23 &146s& 17.2s  \\
  \hline
  \mr{2}{$385\times385\times97$}  & 3.0 & 14.4 & 20 & 185s & 18 & 137s & 26 & 236s & 67s \\
                                  & 4.0 & 10.8 & 26 & 241s & 24 & 180s & 25 & 226s & 64s \\
  \hline
  \mr{2}{$513\times513\times129$} & 4.0 & 14.4 & 27 & 409s & 22 & 294s & 34 & 505s & 170s \\
                                  & 6.0 & 9.6 & 44 & 674s & 37 & 504s & 32 & 474s & 177s \\
  \hline
 \end{tabular}
\caption{3D Overthrust model: solution performance.  }
\label{tab:Overthrust3D}
\end{table}

The last model that we present is the 3D SEG Overthrust model \cite{aminzadeh19973}, which, similarly to the Marmousi 2 model, includes many reflecting layers. The model appears in Fig. \ref{fig:Overthrust}, and Table \ref{tab:Overthrust3D} summarizes the performance results. At the smaller sizes, the upwind ADR discretization requires the most effort to solve, but it becomes more efficient compared with the other to discretizations when the problem gets larger. The shifted Laplacian approach used for the standard and central ADR discretizations yields comparable counts for both. Again, there is a slight edge to ADR central thanks to Stage 1 in Algorithm \ref{alg:ADRlong}. We expect that at larger scales we will see results which are similar to the results in the 2D Marmousi 2 model.

\section{Conclusions and future work}
In this paper we present a new approach for discretizing and solving the Helmholtz equation with a point source. We reformulate the problem based on the Rytov decomposition of the solution, yielding an eikonal equation for the phase and a complex-valued advection-–diffusion-–reaction equation for the amplitude. We choose the phase based on the travel time of the wave, and compute it based on the factored eikonal equation using the Fast Marching algorithm. The factored version yields a more accurate treatment of the point source for the phase, and a relatively smooth solution for the amplitude. The ADR equation is discretized using second order upwind and central difference discretizations, and the solution of the system is achieved using multigrid. Our approach has two main advantages. First, the majority of the solution of the Helmholtz equation is represented by smooth functions, and hence the reformulated problems is more suitable for multigrid computations. Secondly, the obtained solution is not a first arrival anzatz only---it includes all the information of the wave propagation including reflections and inferences.


Our accuracy results show that for models that do not introduce caustics and reflections, our approach yields more accurate solutions than standard Helmholtz discretization, and in particular, does not introduce a phase error. When reflections and caustics are observed, the accuracy of the ADR discretizations is comparable to the standard one. Our performance results show that the standard and central ADR discretizations are both hard to solve. The ADR system using upwind discretization is solved more efficiently, but is less accurate than the ADR discretization with central difference in the presence of reflections.

Our approach is intriguing for geophysical applications, especially full waveform inversion (FWI) where many solutions of the Helmholtz equations are required for a point source. There, the model is adapted iteratively, starting from a smooth model like in Fig. \ref{fig:LinearVel}, until a detailed model like in Fig. \ref{fig:Marmousi2} is estimated in the final stages. Smooth models are encountered frequently, and hence our approach can be beneficial computationally. More importantly, our ADR discretization does not introduce dispersion errors for the main wave, which may be very hard to overcome in the inversion process in FWI.

Our future research aims at exploring the advantages of the new discretizations in the context of FWI, and to further improve the numerical solver for the Helmholtz equation at larger scales and higher wave number. In addition, we aim to extend the presented approach for a general right hand side, which will allow the use of the whole method as a preconditioner.

\section{Acknowledgement}
The research leading to these results has received funding from the European Union's - Seventh Framework Programme (FP7/2007-2013) under grant agreement number 623212 - MC Multiscale Inversion.

{
\bibliographystyle{wileyj}
\bibliography{HelmholtzForwardV4}

\begin{thebibliography}{10}
\providecommand{\url}[1]{\texttt{#1}}
\providecommand{\urlprefix}{URL }
\expandafter\ifx\csname urlstyle\endcsname\relax
  \providecommand{\doi}[1]{doi:\discretionary{}{}{}#1}\else
  \providecommand{\doi}{doi:\discretionary{}{}{}\begingroup
  \urlstyle{rm}\Url}\fi

\bibitem{pratt1999}
Pratt R. Seismic waveform inversion in the frequency domain, part 1: Theory,
  and verification in a physical scale model. \emph{Geophysics}  1999;
  \textbf{64}:888--901.

\bibitem{EpanomeritakisAkcelikGhattasBielak2008}
Epanomeritakis I, Akcelik V, Ghattas O, Bielak J. A {Newton-CG} method for
  large-scale three-dimensional elastic full-waveform seismic inversion.
  \emph{Inverse Problems}  2008; .

\bibitem{virieux2009overview}
Virieux J, Operto S. An overview of full-waveform inversion in exploration
  geophysics. \emph{Geophysics}  2009; \textbf{74}(6):WCC1--WCC26.

\bibitem{krebs09ffw}
Krebs JR, Anderson JE, Hinkley D, Neelamani R, Lee S, Baumstein A, Lacasse MD.
  Fast full-wavefield seismic inversion using encoded sources.
  \emph{Geophysics}  2009; \textbf{74}(6):WCC177--WCC188.

\bibitem{biondi2014simultaneous}
Biondi B, Almomin A. Simultaneous inversion of full data bandwidth by
  tomographic full-waveform inversion. \emph{Geophysics}  2014;
  \textbf{79}(3):WA129--WA140.

\bibitem{metivier2013full}
M{\'e}tivier L, Brossier R, Virieux J, Operto S. Full waveform inversion and
  the truncated newton method. \emph{SIAM Journal on Scientific Computing}
  2013; \textbf{35}(2):B401--B437.

\bibitem{JointEikFWI17}
Treister E, Haber E. Full waveform inversion guided by travel time tomography.
  \emph{SIAM J. Sci. Comput.}  2017; \textbf{39}(5):S587--–S609.

\bibitem{singer2004perfectly}
Singer I, Turkel E. A perfectly matched layer for the helmholtz equation in a
  semi-infinite strip. \emph{Journal of Computational Physics}  2004;
  \textbf{201}(2):439--465.

\bibitem{engquist1979radiation}
Engquist B, Majda A. Radiation boundary conditions for acoustic and elastic
  wave calculations. \emph{Communications on pure and applied mathematics}
  1979; \textbf{32}(3):313--357.

\bibitem{liao1996multifrequency}
Liao Q, McMechan GA. Multifrequency viscoacoustic modeling and inversion.
  \emph{Geophysics}  1996; \textbf{61}(5):1371--1378.

\bibitem{erlangga2006novel}
Erlangga YA, Oosterlee CW, Vuik C. A novel multigrid based preconditioner for
  heterogeneous {Helmholtz} problems. \emph{SIAM Journal on Scientific
  Computing}  2006; \textbf{27}(4):1471--1492.

\bibitem{calandra2013improved}
Calandra H, Gratton S, Pinel X, Vasseur X. An improved two-grid preconditioner
  for the solution of three-dimensional {Helmholtz} problems in heterogeneous
  media. \emph{Numerical Linear Algebra with Applications}  2013;
  \textbf{20}(4):663--688.

\bibitem{poulson2013parallel}
Poulson J, Engquist B, Li S, Ying L. A parallel sweeping preconditioner for
  heterogeneous {3D} {Helmholtz} equations. \emph{SIAM Journal on Scientific
  Computing}  2013; \textbf{35}(3):C194--C212.

\bibitem{haber2011fast}
Haber E, MacLachlan S. A fast method for the solution of the {Helmholtz}
  equation. \emph{Journal of Computational Physics}  2011;
  \textbf{230}(12):4403--4418.

\bibitem{oosterlee2010shifted}
Oosterlee C, Vuik C, Mulder W, Plessix RE. Shifted-laplacian preconditioners
  for heterogeneous {Helmholtz} problems. \emph{Advanced Computational Methods
  in Science and Engineering}. Springer, 2010; 21--46.

\bibitem{airaksinen2007algebraic}
Airaksinen T, Heikkola E, Pennanen A, Toivanen J. An algebraic multigrid based
  shifted-laplacian preconditioner for the {Helmholtz} equation. \emph{Journal
  of Computational Physics}  2007; \textbf{226}(1):1196--1210.

\bibitem{erlangga2008multilevel}
Erlangga YA, Nabben R. On a multilevel {Krylov} method for the {Helmholtz}
  equation preconditioned by shifted laplacian. \emph{Electronic Transactions
  on Numerical Analysis}  2008; \textbf{31}(403-424):3.

\bibitem{cools2014new}
Cools S, Reps B, Vanroose W. A new level-dependent coarse grid correction
  scheme for indefinite {Helmholtz} problems. \emph{Numerical Linear Algebra
  with Applications}  2014; \textbf{21}(4):513--533.

\bibitem{tsuji2015augmented}
Tsuji P, Tuminaro R. Augmented amg-shifted laplacian preconditioners for
  indefinite helmholtz problems. \emph{Numerical Linear Algebra with
  Applications}  2015; \textbf{22}(6):1077--1101.

\bibitem{Tobias2017}
Reps B, Weinzierl T. Complex additive geometric multilevel solvers for
  helmholtz equations on spacetrees. \emph{ACM Transactions on Mathematical
  Software (TOMS)}  2017; \textbf{44}(1):2.

\bibitem{ganesh2017efficient}
Ganesh M, Morgenstern C. An efficient multigrid algorithm for heterogeneous
  acoustic media sign-indefinite high-order fem models. \emph{Numerical Linear
  Algebra with Applications}  2017; \textbf{24}(3).

\bibitem{engquist2011sweeping}
Engquist B, Ying L. Sweeping preconditioner for the {Helmholtz} equation:
  hierarchical matrix representation. \emph{Communications on pure and applied
  mathematics}  2011; \textbf{64}(5):697--735.

\bibitem{van2012preconditioning}
van Leeuwen T, Gordon D, Gordon R, Herrmann FJ. Preconditioning the {Helmholtz}
  equation via row-projections. \emph{74th EAGE Conference \& Exhibition},
  2012.

\bibitem{gordon2013robust}
Gordon D, Gordon R. Robust and highly scalable parallel solution of the
  {Helmholtz} equation with large wave numbers. \emph{Journal of Computational
  and Applied Mathematics}  2013; \textbf{237}(1):182--196.

\bibitem{brandt1997wave}
Brandt A, Livshits I. Wave-ray multigrid method for standing wave equations.
  \emph{Electron. Trans. Numer. Anal}  1997; \textbf{6}:162--181.

\bibitem{olson2010smoothed}
Olson LN, Schroder JB. Smoothed aggregation for {Helmholtz} problems.
  \emph{Numerical Linear Algebra with Applications}  2010;
  \textbf{17}(2-3):361--386.

\bibitem{livshits2014scalable}
Livshits I. A scalable multigrid method for solving indefinite {Helmholtz}
  equations with constant wave numbers. \emph{Numerical Linear Algebra with
  Applications}  2014; \textbf{21}(2):177--193.

\bibitem{leung2007eulerian}
Leung S, Qian J, Burridge R. Eulerian {Gaussian} beams for high-frequency wave
  propagation. \emph{Geophysics}  2007; \textbf{72}(5):SM61--SM76.

\bibitem{luo2011factored}
Luo S, Qian J. Factored singularities and high-order lax--friedrichs sweeping
  schemes for point-source traveltimes and amplitudes. \emph{Journal of
  Computational Physics}  2011; \textbf{230}(12):4742--4755.

\bibitem{luo2012higher}
Luo S, Qian J, Zhao H. Higher-order schemes for {3D} first-arrival traveltimes
  and amplitudes. \emph{Geophysics}  2012; \textbf{77}(2):T47--T56.

\bibitem{luo2014fast}
Luo S, Qian J, Burridge R. Fast {Huygens} sweeping methods for {Helmholtz}
  equations in inhomogeneous media in the high frequency regime. \emph{Journal
  of Computational Physics}  2014; \textbf{270}:378--401.

\bibitem{crandall1983viscosity}
Crandall MG, Lions PL. Viscosity solutions of {Hamilton}-{Jacobi} equations.
  \emph{Transactions of the American Mathematical Society}  1983;
  \textbf{277}(1):1--42.

\bibitem{rouy1992viscosity}
Rouy E, Tourin A. A viscosity solutions approach to shape-from-shading.
  \emph{SIAM Journal on Numerical Analysis}  1992; \textbf{29}(3):867--884.

\bibitem{tsitsiklis1995efficient}
Tsitsiklis JN. Efficient algorithms for globally optimal trajectories.
  \emph{Automatic Control, IEEE Transactions on}  1995;
  \textbf{40}(9):1528--1538.

\bibitem{sethian1996fast}
Sethian JA. A fast marching level set method for monotonically advancing
  fronts. \emph{Proceedings of the National Academy of Sciences}  1996;
  \textbf{93}(4):1591--1595.

\bibitem{sethian1999fast}
Sethian JA. Fast marching methods. \emph{SIAM review}  1999;
  \textbf{41}(2):199--235.

\bibitem{tsai2003fast}
Tsai YHR, Cheng LT, Osher S, Zhao HK. Fast sweeping algorithms for a class of
  {Hamilton}-{Jacobi} equations. \emph{SIAM journal on numerical analysis}
  2003; \textbf{41}(2):673--694.

\bibitem{zhao2005fast}
Zhao H. A fast sweeping method for eikonal equations. \emph{Mathematics of
  computation}  2005; \textbf{74}(250):603--627.

\bibitem{zhang2006high}
Zhang YT, Zhao HK, Qian J. High order fast sweeping methods for static
  {Hamilton}-{Jacobi} equations. \emph{Journal of Scientific Computing}  2006;
  \textbf{29}(1):25--56.

\bibitem{kao2004lax}
Kao CY, Osher S, Qian J. Lax--friedrichs sweeping scheme for static
  {Hamilton}-{Jacobi} equations. \emph{Journal of Computational Physics}  2004;
  \textbf{196}(1):367--391.

\bibitem{qian2007fast}
Qian J, Zhang YT, Zhao HK. A fast sweeping method for static convex
  {Hamilton}-{Jacobi} equations. \emph{Journal of Scientific Computing}  2007;
  \textbf{31}(1):237--271.

\bibitem{fomel2009fast}
Fomel S, Luo S, Zhao H. Fast sweeping method for the factored eikonal equation.
  \emph{Journal of Computational Physics}  2009; \textbf{228}(17):6440--6455.

\bibitem{pica1997fast}
Pica A, \emph{et~al.}. Fast and accurate finite-difference solutions of the
  {3D} eikonal equation parametrized in celerity. \emph{67th Ann. Internat.
  Mtg, Soc. of Expl. Geophys}  1997; :1774--1777.

\bibitem{TH2016}
Treister E, Haber E. A fast marching algorithm for the factored eikonal
  equation. \emph{Journal of Computational Physics}  2016;
  \textbf{324}:210--225.

\bibitem{luo2012fast}
Luo S, Qian J. Fast sweeping methods for factored anisotropic eikonal
  equations: multiplicative and additive factors. \emph{Journal of Scientific
  Computing}  2012; \textbf{52}(2):360--382.

\bibitem{LouQianBurridge2014}
Luo S, Qian J, Burridge R. High-order factorization based high-order hybrid
  fast sweeping methods for point-source eikonal equations. \emph{SIAM Journal
  on Numerical Analysis}  2014; \textbf{52}(1):23--44, \doi{10.1137/120901696}.
  \urlprefix\url{http://dx.doi.org/10.1137/120901696}.

\bibitem{noble2014accurate}
Noble M, Gesret A, Belayouni N. Accurate 3-d finite difference computation of
  traveltimes in strongly heterogeneous media. \emph{Geophysical Journal
  International}  2014; \textbf{199}(3):1572--1585.

\bibitem{singer1998high}
Singer I, Turkel E. High-order finite difference methods for the helmholtz
  equation. \emph{Computer Methods in Applied Mechanics and Engineering}  1998;
  \textbf{163}(1-4):343--358.

\bibitem{singer2006sixth}
Singer I, Turkel E. Sixth-order accurate finite difference schemes for the
  helmholtz equation. \emph{Journal of Computational Acoustics}  2006;
  \textbf{14}(03):339--351.

\bibitem{operto20073d}
Operto S, Virieux J, Amestoy P, L’Excellent JY, Giraud L, Ali HBH. 3d
  finite-difference frequency-domain modeling of visco-acoustic wave
  propagation using a massively parallel direct solver: A feasibility study.
  \emph{Geophysics}  2007; \textbf{72}(5):SM195--SM211.

\bibitem{turkel2013compact}
Turkel E, Gordon D, Gordon R, Tsynkov S. Compact 2d and 3d sixth order schemes
  for the helmholtz equation with variable wave number. \emph{Journal of
  Computational Physics}  2013; \textbf{232}(1):272--287.

\bibitem{bayliss1985accuracy}
Bayliss A, Goldstein CI, Turkel E. On accuracy conditions for the numerical
  computation of waves. \emph{Journal of Computational Physics}  1985;
  \textbf{59}(3):396--404.

\bibitem{BHM00}
Briggs WL, Henson VE, McCormick SF. \emph{A multigrid tutorial}. Second edn.,
  SIAM, 2000.

\bibitem{TOS01}
Trottenberg U, Oosterlee C, Sch\"{u}ller A. \emph{Multigrid}. Academic Press:
  London and San Diego, 2001.

\bibitem{Yav06}
Yavneh I. Why multigrid methods are so efficient. \emph{IEEE: Computing in
  Science and Engineering}  2006; \textbf{8}(6):12--22.

\bibitem{knibbe2011gpu}
Knibbe H, Oosterlee CW, Vuik C. {GPU} implementation of a {Helmholtz} {Krylov}
  solver preconditioned by a shifted laplace multigrid method. \emph{Journal of
  Computational and Applied Mathematics}  2011; \textbf{236}(3):281--293.

\bibitem{saad1993flexible}
Saad Y. A flexible inner-outer preconditioned {GMRES} algorithm. \emph{SIAM
  Journal on Scientific Computing}  1993; \textbf{14}(2):461--469.

\bibitem{elman2001multigrid}
Elman HC, Ernst OG, O'leary DP. A multigrid method enhanced by krylov subspace
  iteration for discrete helmholtz equations. \emph{SIAM Journal on scientific
  computing}  2001; \textbf{23}(4):1291--1315.

\bibitem{jInv17}
Ruthotto L, Treister E, Haber E. {jInv} -- a flexible {Julia} package for {PDE}
  parameter estimation. \emph{SIAM J. Sci. Comput.}  2017;
  \textbf{39}(5):S702–--S722.

\bibitem{Julia}
Bezanson J, Edelman A, Karpinski S, Shah VB. Julia: A fresh approach to
  numerical computing. \emph{SIAM Review}  2017; \textbf{59}(1):65--98,
  \doi{10.1137/141000671}.
  \urlprefix\url{http://julialang.org/publications/julia-fresh-approach-BEKS.pdf}.

\bibitem{martin2006marmousi2}
Martin GS, Wiley R, Marfurt KJ. Marmousi2: An elastic upgrade for marmousi.
  \emph{The Leading Edge}  2006; \textbf{25}(2):156--166.

\bibitem{notay2008recursive}
Notay Y, Vassilevski PS. Recursive {Krylov}-based multigrid cycles.
  \emph{Numerical Linear Algebra with Applications}  2008;
  \textbf{15}(5):473--487.

\bibitem{aminzadeh19973}
Aminzadeh F, Jean B, Kunz T. \emph{3-D salt and overthrust models}. Society of
  Exploration Geophysicists, 1997.

\end{thebibliography}
}
\end{document}